\documentclass[prb, aps, twocolumn, floatfix, superscriptaddress]{revtex4-2}
\usepackage{amsmath}
\usepackage{amssymb}
\usepackage{graphicx}
\usepackage{color}

\newcommand{\be}{\begin{equation}}
\newcommand{\ee}{\end{equation}}
\newcommand{\bea}{\begin{eqnarray}}
\newcommand{\eea}{\end{eqnarray}}
\newcommand{\bse}{\begin{subequations}}
\newcommand{\ese}{\end{subequations}}
\newcommand{\emb}{${\rm EuMg_2Bi_2}$}
\newcommand{\ems}{${\rm EuMg_2Sb_2}$}
\newcommand{\cas}{${\rm CaAl_2Si_2}$}
\newcommand{\esa}{${\rm EuSn_2As_2}$}

\begin{document}

\title{A-type antiferromagnetic order in semiconducting ${\rm EuMg_2Sb_2}$ single crystals}

\author{Santanu Pakhira}
\affiliation{Ames Laboratory, Iowa State University, Ames, Iowa 50011, USA}
\author{Farhan Islam}
\affiliation{Ames Laboratory, Iowa State University, Ames, Iowa 50011, USA}
\affiliation{Department of Physics and Astronomy, Iowa State University, Ames, Iowa 50011, USA}
\author{Evan O'Leary}
\affiliation{Ames Laboratory, Iowa State University, Ames, Iowa 50011, USA}
\affiliation{Department of Physics and Astronomy, Iowa State University, Ames, Iowa 50011, USA}
\author{M. A. Tanatar}
\affiliation{Ames Laboratory, Iowa State University, Ames, Iowa 50011, USA}
\affiliation{Department of Physics and Astronomy, Iowa State University, Ames, Iowa 50011, USA}
\author{Thomas Heitmann}
\affiliation{The Missouri Research Reactor and Department of Physics and Astronomy, University of Missouri, Columbia, Missouri 65211, USA}
\author{R. Prozorov}
\affiliation{Ames Laboratory, Iowa State University, Ames, Iowa 50011, USA}
\affiliation{Department of Physics and Astronomy, Iowa State University, Ames, Iowa 50011, USA}
\author{Adam Kaminski}
\affiliation{Ames Laboratory, Iowa State University, Ames, Iowa 50011, USA}
\affiliation{Department of Physics and Astronomy, Iowa State University, Ames, Iowa 50011, USA}
\author{David Vaknin}
\affiliation{Ames Laboratory, Iowa State University, Ames, Iowa 50011, USA}
\affiliation{Department of Physics and Astronomy, Iowa State University, Ames, Iowa 50011, USA}
\author{D. C. Johnston}
\affiliation{Ames Laboratory, Iowa State University, Ames, Iowa 50011, USA}
\affiliation{Department of Physics and Astronomy, Iowa State University, Ames, Iowa 50011, USA}

\date{\today}

\begin{abstract}

Eu-based Zintl-phase materials Eu$A_2Pn_2$ ($A$ = Mg, In, Cd, Zn; $Pn$ = Bi, Sb, As, P) have generated significant recent interest owing to the complex interplay of magnetism and band topology. Here, we investigated the electronic, magnetic, and electronic properties of the layered Zintl-phase single crystals of \ems\ with the trigonal \cas\ crystal structure (space group $P\overline{3}m1$). Electrical resistivity measurements complemented with angle-resolved photoemission spectroscopy (ARPES) studies find an activated behavior with the intrinsic conductivity at high temperatures indicating a semiconducting electronic ground state with a narrow energy gap of 370~meV\@. Magnetic susceptibility and zero-field heat-capacity measurements indicate that the compound undergoes antiferromagnetic (AFM) ordering at the N\'{e}el temperature $T_{\rm N} = 8.0(2)$~K\@.  Zero-field neutron-diffraction measurements reveal that the AFM ordering is A-type where the Eu ordered moments (Eu$^{2+},\, S= 7/2$) arranged in $ab$-plane layers are aligned ferromagnetically in the $ab$ plane with the Eu moments in adjacent layers aligned antiferromagnetically. We also find that Eu-moment reorientation in the trigonal AFM domains within the $ab$ planes occurs below below $T_{\rm N}$ at low fields $<0.05$~T due to very small in-plane anisotropy. Although isostructural semimetallic \emb\ is reported to host Dirac surface states, the observation of narrow-gap semiconducting behavior in \ems\ implies a strong role of spin-orbit coupling in tuning the electronic states of these materials.

\end{abstract}

\maketitle

\section{Introduction}

Tuning of electronic band structure through coupling between lattice, charge, and electronic degrees of freedom is key to recent discoveries in condensed-matter physics and material science. Materials with nontrivial band topology have been extensively studied owing to their possible applications in dissipationless electronic transport~\cite{Hasan2010, Qi2011, Yan2017, Tokura2019}. Many rare-earth-based magnetic materials also belong to this category and have been reported to host novel electronic states through a complex interplay of magnetism, spin-orbit coupling (SOC), and band topology~\cite{Hirschberger2016, Shekhar2018, Borisenko2019, Li2019}. For materials on the verge of time-reversal-symmetry breaking associated with magnetism and the presence or absence of SOC, the electronic states of these materials can be tuned between metallic, insulating, semimetallic, narrow-gap and wide-gap semiconducting behavior.

In recent years, studies of Eu-based intermetallic compounds carried out in the search for novel electronic states have been reported~\cite{Masuda2016, Soh2019, Jo2020, Riberolles2021}.  \emb\ is one such material which belongs to a class of rare-earth-based compounds that exhibit novel electronic states arising from a complex interplay of magnetism and electron-band topology~\cite{kabir2019, Pakhira2020, Marshall2020}. It has recently gained interest because it possesses Dirac points located at different energies with respect to the Fermi energy~\cite{kabir2019}.  EuMg$_2Pn_2$ ($Pn = $ P, As, Sb, or Bi) crystallize in the trigonal CaAl$_2$Si$_2$ crystal structure with space group $P\overline{3}m1$, (No. 164)~\cite{May2011}. The Eu atoms form a triangular lattice in the $ab$~plane and these planes are stacked along the $c$~axis. The Mg and $Pn$ atoms are arranged in two triangular layers between adjacent layers of Eu atoms. \emb\ is a semimetal and exhibits A-type antiferromagnetic (AFM) order below the  N\'eel temperature $T_{\rm N} = 6.7$~K\@~\cite{Pakhira2020, Pakhira2021}. In this case the Eu moments within an $a$~plane are ferromagnetically aligned in the $ab$~plane, where the moments in adjacent Eu planes along the $c$~axis are aligned antiferromagnetically~\cite{Pakhira2021}. Moment reorientation within the $ab$~plane associated with weak in-plane anisotropy was also observed at low fields below $T_{\rm N}$~\cite{Pakhira2021}. Recently, it was shown that substituting Ca for Eu significantly affects the electronic states where a semimetal-to-semiconductor transition occurs~\cite{Marshall2022}. This indicates that EuMg$_2Pn_2$ compounds offer a fertile ground to study the tunability of the electronic states in these materials.

\ems\ is isostructural to \emb.  Polycrystalline \ems\ was reported to exhibit AFM ordering below $T_{\rm N} = 8.2(3)$~K based on magnetic susceptibility~$\chi$ measurements in the magnetic field $H = 0.002$~T\@~\cite{Wartenberg2002}. No information about single-crystal growth, the nature of the AFM ordering, it's field evolution, or its electronic properties is known so far to our knowledge. Since the SOC is considerably smaller in Sb compared to Bi, it is interesting to study the magnetic and electronic properties of \ems\ in order to probe the role of SOC in these materials on the associated electronic and magnetic states.

Here, we report the growth of \ems\ single crystals and their crystallographic, electronic, and magnetic properties studied by means of room-temperature x-ray powder diffraction (XRD), temperature~$T$-dependent electrical resistivity~$\rho$, heat capacity $C_{\rm p}$, $\chi(H,T)$ and magnetization $M(H)$ isotherm measurements.  We also report the results of angle-resolved photoemission spectroscopy (ARPES) and zero-field neutron-diffraction measurements. In contrast to the semimetallic behavior observed in isostructural \emb, \ems\ is found to be a narrow-gap semiconductor, as revealed by our $\rho(T)$ and ARPES measurements. We suggest that the reduction in SOC by introducing Sb in place of Bi might be responsible for the change in the electronic band structure. Our $\chi(T)$ and $C_{\rm p}(T)$ data reveal that \ems\ undergoes a long-range AFM transition below $T_{\rm N} = 8.0(2)$~K with the Eu$^{2+}$ moments with spin $S=7/2$ aligned in the $ab$~plane. The zero-field neutron measurements show the AFM ordering to be A-type below $T_{\rm N}$. An additional cusp in the in-plane $\chi(T)$ of unknown origin is observed at $T \approx 3$~K for low applied magnetic fields.  A reorientation of the in-plane ordered-moment alignment is observed in low $ab$-plane magnetic fields via $M(H,T)$ measurements, indicating a very small in-plane anisotropy energy.

The experimental details are given in the following Sec.~\ref{Sec:ExpDet}.  The results and discussion of the various measurements are presented in Sec.~\ref{Sec:Results}, and a summary is provided in Sec.~\ref{Sec:Conclu}.  The dependence of the Cartesian eigenvalues and eigenvectors of the magnetic-dipole-interaction tensor versus the $c/a$ ratio from 0.5 to 3 in 0.1 increments for a stacked simple-hexagonal spin lattice with collinear A-type AFM order and with the moments aligned in the $ab$~plane is given in both graphical and tabular forms in the Appendix.

\section{\label{Sec:ExpDet} Experimental Details}

\ems\ single crystals were grown using the self-flux method with starting composition Eu:Mg:Sb = 1:4:16, where the self-flux therefore had composition Mg$_2$Sb$_{14}$. The elements were loaded in an alumina crucible with a quartz wool filter and sealed inside a silica tube under $1/4$ atm high-purity argon. The assembly was then heated to 900~$^{\circ}$C at a rate 50~$^{\circ}$C/h followed by a dwell of 12~h at that temperature. It was then cooled to 750~$^{\circ}$C at a rate of 3~$^{\circ}$C/h. The crystals were obtained by removing the flux through centrifugation at that temperature. Structural characterization was performed by room-temperature powder x-ray diffraction (XRD) measurements on ground crystals using a Rigaku Geigerflex x-ray diffractometer with Cu-$K_\alpha$ radiation. Crystal-structure analysis was performed using Rietveld refinement with the FULLPROF software package~\cite{Carvajal1993}. The sample homogeneity and chemical composition were confirmed using a JEOL scanning-electron microscope (SEM) equipped with an energy-dispersive x-ray spectroscopy (EDS) analyzer. The magnetic measurements were carried out in a Magnetic-Properties-Measurement System (MPMS) from Quantum Design, Inc., in the $T$ range 1.8--300 K and with $H$ up to 5.5 T (1~T~$\equiv10^4$~Oe). A Physical Properties Measurement System (PPMS, Quantum Design, Inc.) was used to measure $C_{\rm p}(T)$ and $\rho(T)$ in the $T$ range 1.8--300~K\@.

The samples used for four-probe in-plane $\rho(T)$ measurements were as-grown single crystals with natural facets and with typical dimensions \mbox{(2--3)\,$\times 1 \times$\,(0.5--1)~mm$^3$}. Because of high reactivity of the crystals with air we were not able to shape the crystals into resistivity bars. The longer side of the sample was along an arbitrary direction in the hexagonal crystal plane. Contacts to fresh surfaces of the crystals were made by attaching 50~$\mu$m diameter silver wires with In solder~\cite{Tanatar2016}. Because of the irregular crystal shape there is a large uncertainty associated with determination of the geometric factor.

ARPES data were collected using an ARPES spectrometer that consists of a Scienta R8000 electron analyzer and electron-cyclotron-resonance helium discharge lamp by Gamma Data with custom-designed focusing optics. We used 21.2 eV photon energy from the He-I line. The angular resolution was set at $\sim 0.1^{\circ}$ and $1^{\circ}$ along and perpendicular to the direction of the analyzer slit, respectively, and the energy resolution was set at 10 meV\@. Samples were cleaved \textit{in-situ} at a base pressure lower than 2$\times$ 10$^{-11}$ Torr and a temperature of 11~K\@. The samples were kept at the cleaving temperature during measurements.

Single-crystal neutron-diffraction experiments were performed in zero applied magnetic field using the TRIAX triple-axis spectrometer at the University of Missouri Research Reactor (MURR). An incident neutron beam with energy 14.7~meV was directed at the sample using a pyrolytic graphite (PG) monochromator. A PG analyzer was used to reduce the background. The neutron wavelength harmonics were removed from the beam using PG filters placed before the monochromator and in between the sample and analyzer. The beam divergence was limited using collimators before the monochromator; between the monochromator and sample; sample and analyzer; and between the analyzer and detector of $60^\prime-60^\prime-40^\prime-40^\prime$, respectively.  A 30~mg \ems\ crystal was mounted on the cold tip of an Advanced Research Systems closed-cycle refrigerator with a nominal base temperature of 4~K (in this study 6.6 K was the lowest temperature achieved). The crystal was aligned in the $(H0L)$  scattering plane. The lattice parameters at 6.6~K were determined to be $a=4.6531(5)$~{\AA} and $c = 7.6668(5)$~{\AA}.

\section{\label{Sec:Results} Results and Discussion}

\subsection{\label{Sec:structure} X-ray diffraction and crystal structure}

\begin{figure}
\includegraphics [width=3in]{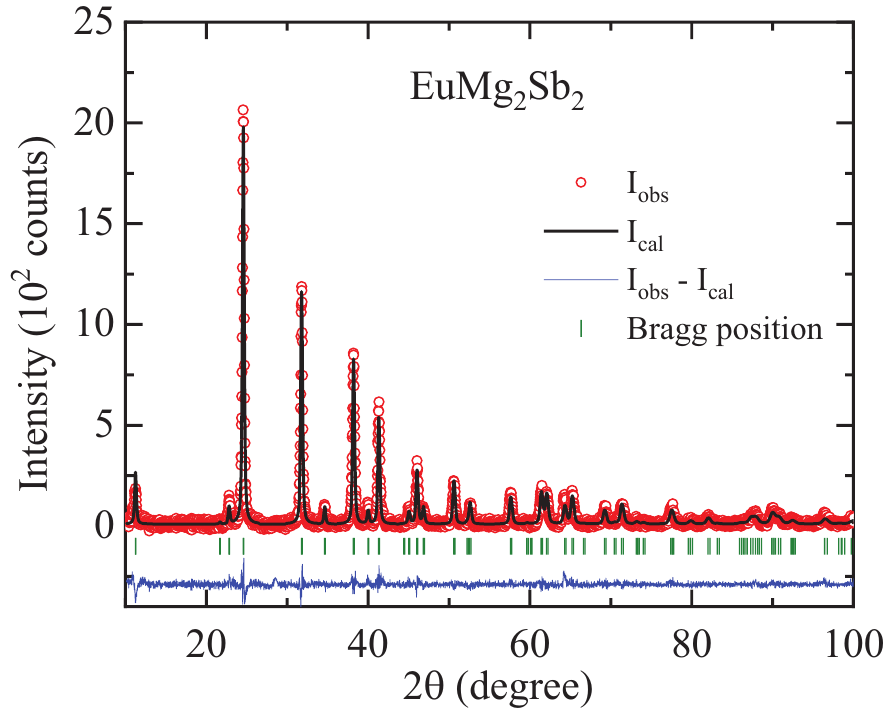}
\caption {Room-temperature powder x-ray diffraction (XRD) patterns of crushed single crystals of \ems\ along with the Rietveld refinement. The open red circles are the experimental data points, the black solid line is the refined pattern, the difference between the experimental and refined diffraction patterns is shown by blue solid curve, and the allowed Bragg positions are marked by the green vertical bars.}
\label{Fig_XRD}
\end{figure}

The room-temperature powder x-ray diffraction (XRD) pattern of the crushed \ems\ single crystals is shown in Fig.~\ref{Fig_XRD}. All the peaks were indexed with the \cas-type crystal structure with space group $P\overline{3}m1$. The refined lattice parameters are $a = b = 4.6861(3)$~\AA\ and $c = 7.7231(5)$~\AA, consistent with the previous results for polycrystalline material~\cite{Wartenberg2002}. The composition obtained from the Rietveld refinement of the XRD data is EuMg$_{1.99(2)}$Sb$_{2.02(3)}$. The SEM-EDX measurements carried out at multiple points on the surfaces of the crystals further confirm the homogeneity and composition EuMg$_{1.96(4)}$Sb$_{2.01(3)}$.  Both compositions are thus consistent within the errors with the stoichiometric composition EuMg$_2$Sb$_2$.

\subsection{\label{Sec:Resis} Electrical resistivity}

\begin{figure}
\includegraphics [width=8cm]{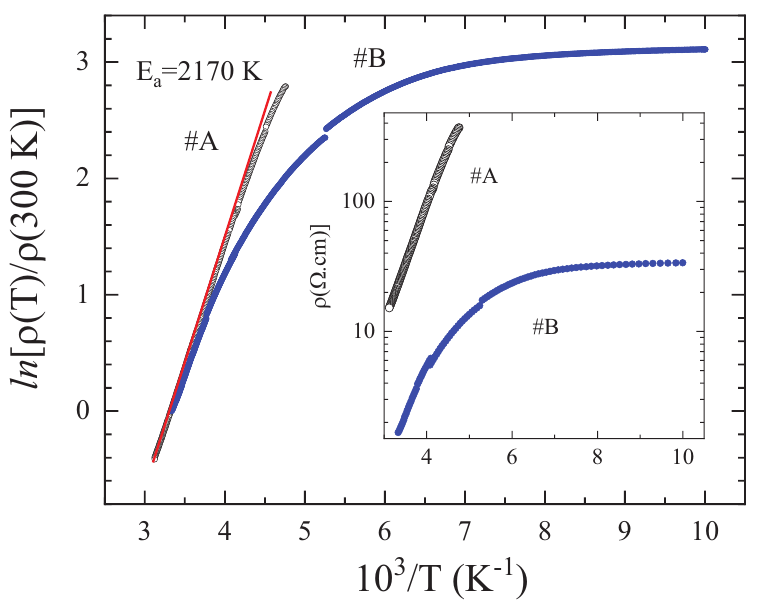}
\caption {Electrical resistivity of EuMg$_2$Sb$_2$ as determined from measurements on two crystals \#A and \#B\@. The data are presented using an Arrhenius plot. The inset shows the measured resistivity data, whereas the main panel presents the normalized resistivity $\ln[\rho(T)/\rho(300$~K)] to remove the large uncertainty in the geometric factor. The two samples show a similar intrinsic slope (activation energy)  $E_{\rm a}/k_{\rm B}=2170$~K of the curves at high temperatures, where $k_{\rm B}$ is Boltzmann's constant. The lower resistivity values of sample \#B, together with saturation of the resistivity at higher temperatures, suggest a larger  contribution of the extrinsic conductivity controlled by impurities and/or defects.}
\label{resistivity}
\end{figure}

Measurements of the temperature-dependent electrical resistivity $\rho$ were performed on two different \ems\ crystals \#A and \#B, as shown in Fig.~\ref{resistivity}. The behavior of both is typical of a semiconductor.  The $\rho$(300~K) for sample \#A was determined as 16~$\Omega$-cm, and for sample \#B as 1.5~$\Omega$-cm (see the inset of Fig.~\ref{resistivity}).  Because of the poorly-defined sample geometry, part of this difference comes from the nearly two-fold uncertainty in the geometric factor. The $\rho(T)$ shows activated behavior, and is presented using Arrhenius plots with an inverse temperature scale 1000/$T$\@.  The inset in Fig.~\ref{resistivity} shows the measured resistivity values $\rho(T)$, whereas the main panel shows the natural logarithm of the normalized resistivity $\ln[\rho(T)/\rho(300~\rm K)]$.

As can be seen from the inset, the difference in resistivity values leads to a faster saturation of $\rho$ of sample~\#B resulting from impurity- and/or defect-controlled conduction. This observation suggests that at least part of the difference in $\rho(300~\rm K)$ comes from a difference in the defect/impurity levels in the crystals. On the other hand, comparison of the $\rho(T)$ data for the two crystals using the normalized $\rho/\rho(300~\rm K)$ scale in the main panel of Fig.~\ref{resistivity}, which largely removes the uncertainty of the geometric factor, reveals a very similar slope of the Arrhenius plots at high $T$\@. This finding suggests that the resistivity in this $T$ range is intrinsic, and is controlled by the energy gap in the sample.  As can be seen from the straight red line fit in the main panel in Fig.~\ref{resistivity} the activation energy as determined from the slope of Arrhenius plot is $E_{\rm a}/k_{\rm B}=2170$~K for both crystals, where $k_{\rm B}$ is \mbox{Boltzmann's} constant. For intrinsic conductivity this corresponds to an energy gap of $E_{\rm gap}/k_{\rm B}=2E_{\rm a}/k_{\rm B}=4340$~K or $E_{\rm gap} = 0.3740$~eV\@. For comparison, this value is about two times less that the indirect gap of silicon and is in a range similar to that for different narrow-gap semiconductors reported earlier~\cite{Hadano2009, Fender2021, Piva2021, Takahashi2011}.

\subsection{\label{Sec:ARPES} Angle-resolved photoemission spectroscopy (ARPES)}

\begin{figure}
\includegraphics [width=3.1in]{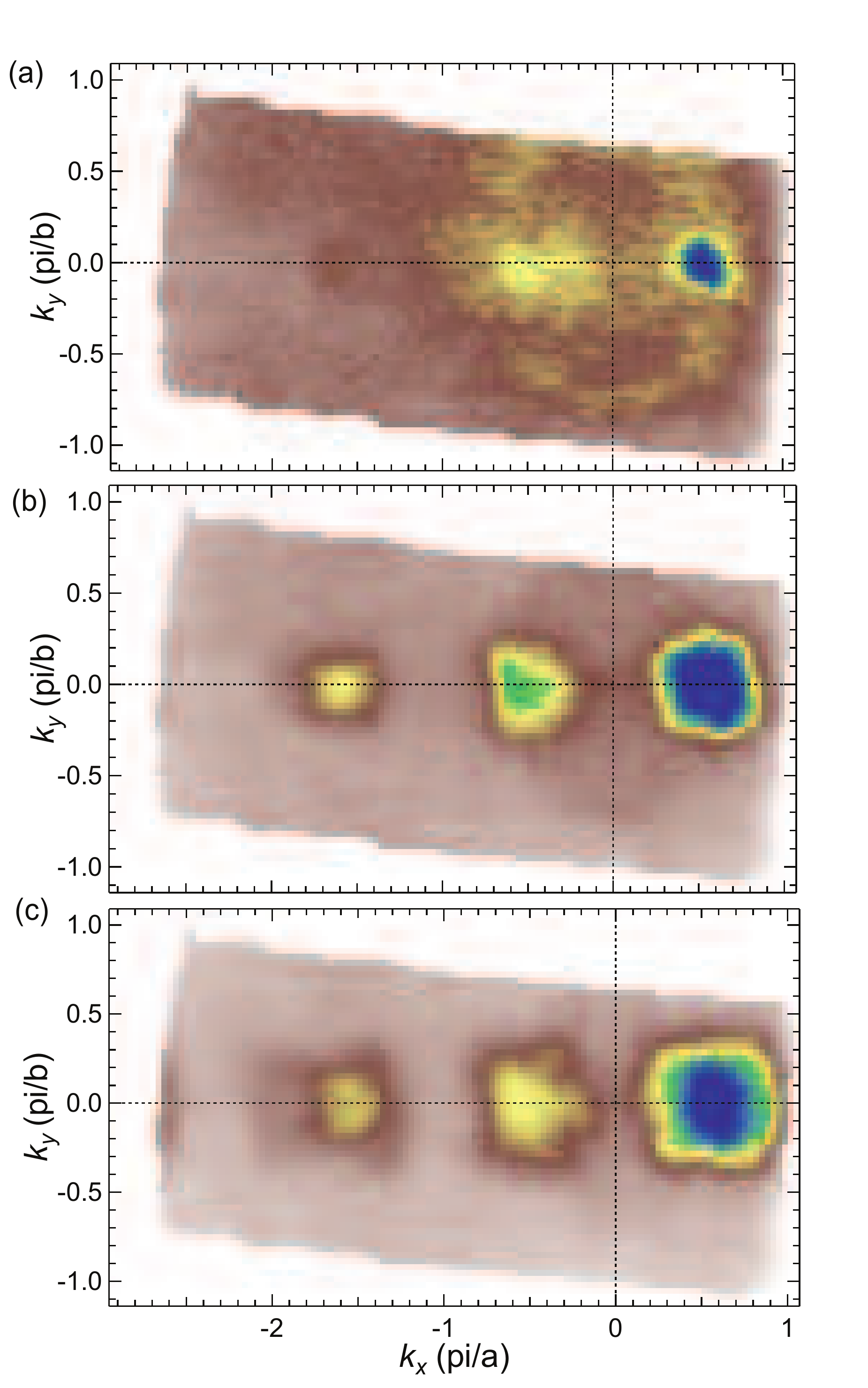}
\caption {Iso-energy cuts of \ems.  (a)~Iso-energy cut at the Fermi level produced by ARPES data, integrated within 10 meV of the Fermi level. (b,c)~Iso-energy cuts taken 0.2 eV and 0.4 eV below the Fermi level, respectively.}
\label{Fig_ARPES_1}
\end{figure}

\begin{figure}
\includegraphics [width=3.5in]{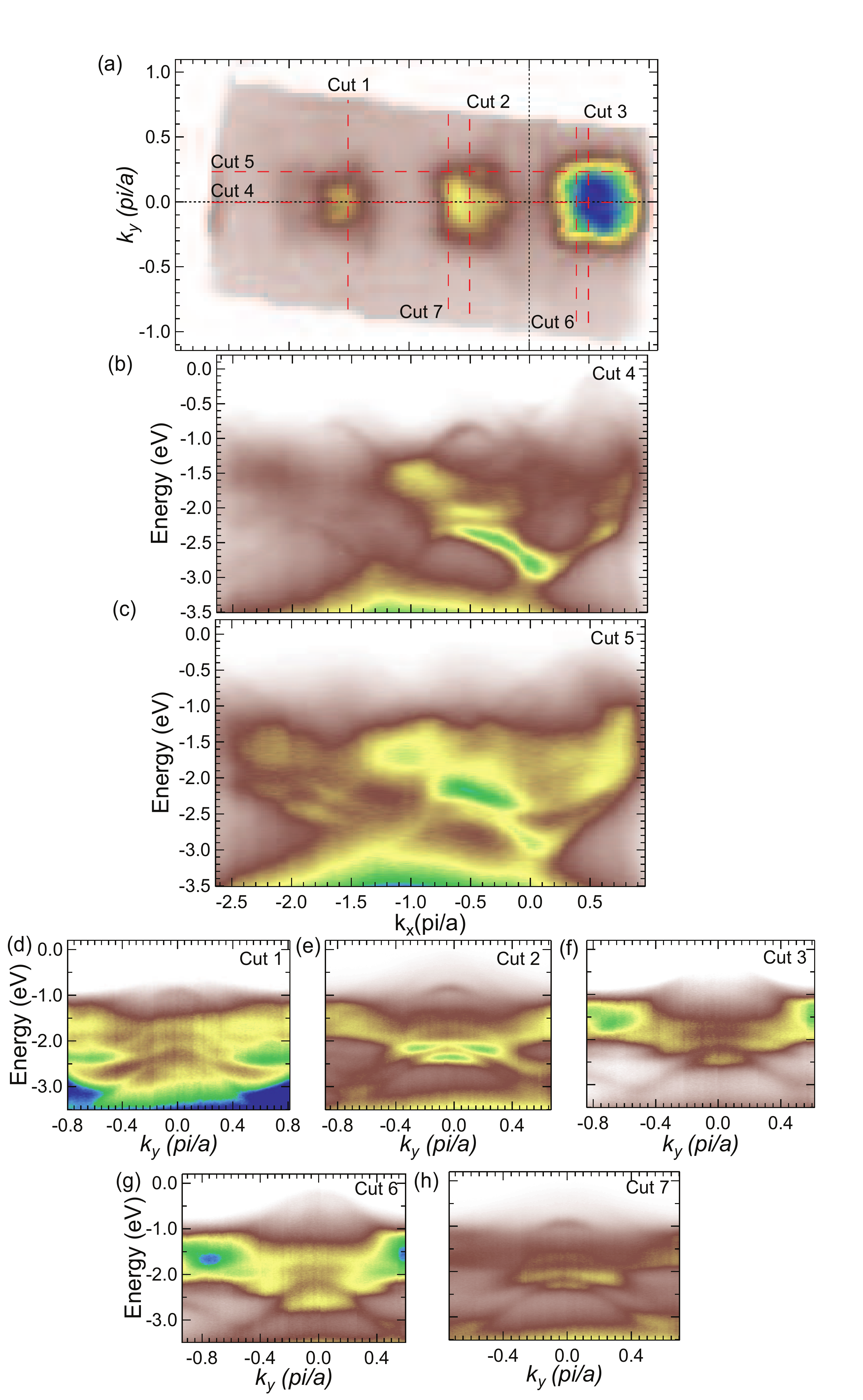}
\caption {Energy dispersion in momentum space for \ems\ (a) Iso-energy cut taken 0.36 eV below the Fermi level. \mbox{(b--h)} Cuts taken in momentum space, cut locations indicated above.}
\label{Fig_ARPES_2}
\end{figure}

The ARPES measurements also predict semiconducting behavior in \ems. There is one band crossing the Fermi level as shown in Fig.~\ref{Fig_ARPES_1}(a). This band has very low occupancy, seen in Fig.~\ref{Fig_ARPES_2}(f), as its intensity is washed out by the intensity of the lower bands, leading to semiconducting behavior. More states open up below the Fermi level, as seen in Figs.~\ref{Fig_ARPES_1}(b,c). Energy cuts through momentum space are shown in Figs.~\ref{Fig_ARPES_2}(a--h). The first energy band with high occupancy is centered around 1.5~eV below the Fermi level as seen in Figs.~\ref{Fig_ARPES_2}(b--h).

\subsection{\label{Sec:Neutron} Zero-field neutron diffraction}

\begin{figure}
\centering
\includegraphics[width=3.4 in]{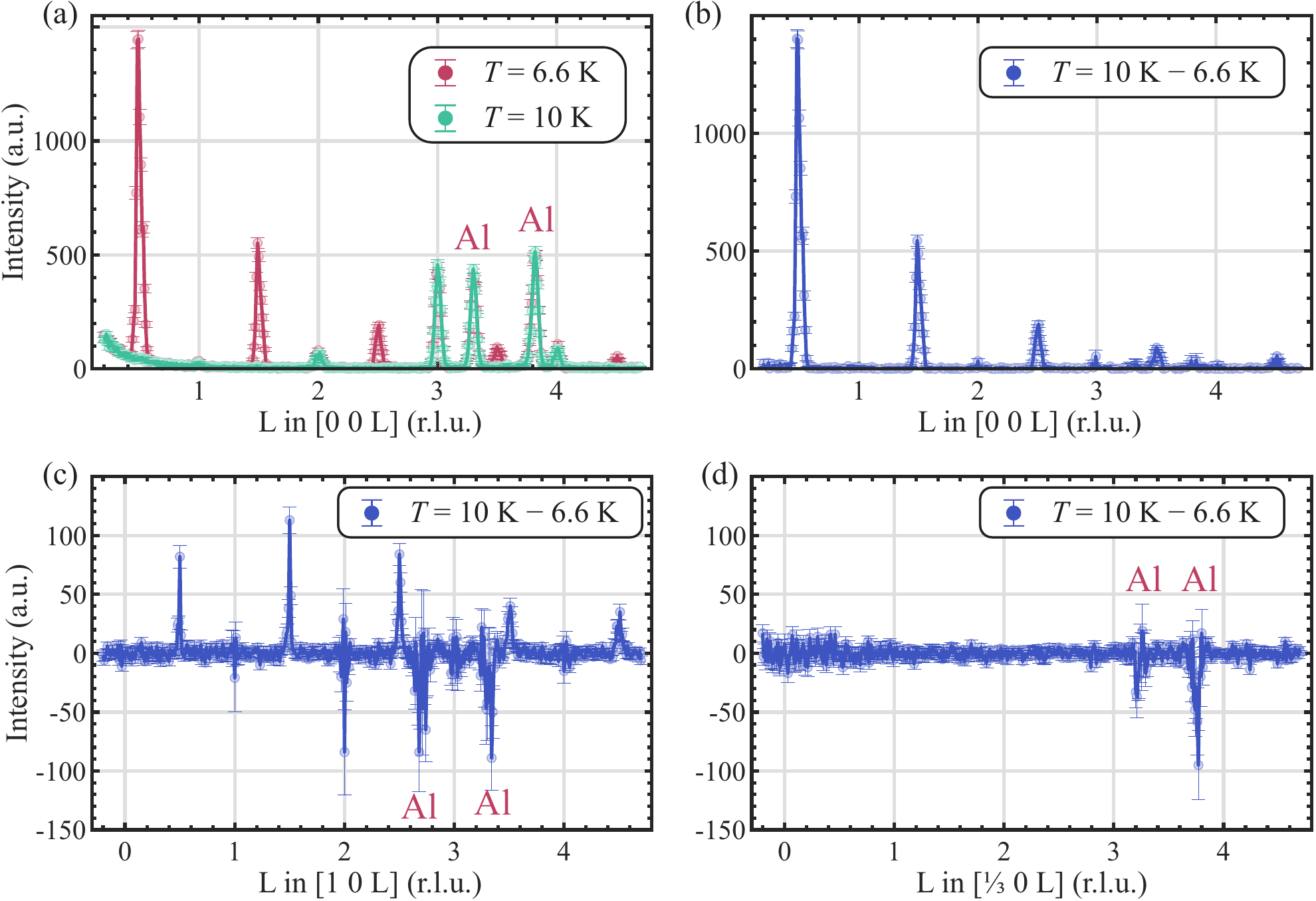}
\caption{(a) Neutron-diffraction pattern along $(00L)$ of single-crystal \ems\ at 6.6 and 10~K as indicated. The aluminum Bragg reflections marked on the figure originate from the sample holder. The magnetic Bragg reflections are obtained by subtracting the diffraction pattern at 10~K from that at 6~K for (b)~$(00L)$, (c)~$(10L)$, and (d)~$(\frac{1}{3}0L)$ scans. The difference patterns in (b--c) show clear magnetic peaks at half-integer $L$ up to $L=4.5$, whereas no such peak is observed in (d) at half integer $L$, consistent with A-type AFM, {\it i.e}, the $H = 0$ ground state is such that the intraplane moments are ferromagnetically aligned in the $ab$ plane while the moments in adjacent Eu planes along the $c$~axis are aligned antiferromagnetically.}
\label{Fig:00l}
\end{figure}

Figure~\ref{Fig:00l}(a) shows neutron-diffraction scans along $(00L)$ (in reciprocal-lattice units) at 6.6 and 10~K, where reflections at half-integer $L$ values are apparent at \mbox{$T = 6.6$~K\@.}  For more clarity, Fig.~\ref{Fig:00l}(b) shows the difference between these two scans, where within experimental uncertainty, there is no evidence for other reflections associated with a modulated structure along the $c$~axis.  Similar differences [i.e., $I{\rm (6.6~K)} - I({\rm 10~K)}]$ for a scan along $(10L)$, shown in Fig.~\ref{Fig:00l}(c), also reveal new peaks at half-integer $L$ values. Qualitatively, these newly-emerging Bragg reflections indicate a doubling of the unit cell along the $c$~axis. We also note that the intensities of the new peaks become weaker at larger $L$ values, roughly following the falloff expected from the magnetic form factor of Eu$^{2+}$. These qualitative observations unequivocally establish that these reflections are associated with A-type AFM ordering with AFM propagation vector $\vec{\tau} = \left(0,0,\frac{1}{2}\right)$, consisting of layers of moments aligned ferromagnetically in the $ab$~plane, with moments in adjacent planes along the $c$~axis aligned antiferromagnetically. Figure \ref{Fig:00l}(d) shows a (1/3,0,$L$) scan with no peaks at half-integer $L$, consistent with ferromagnetic (FM) in-plane ordering. The $\chi(T)$ data discussed in the following section~\ref{Sec:MagSus} also suggest that the ordered moments are aligned in the $ab$~plane.

\begin{figure}
\centering
\includegraphics[width=3.5 in]{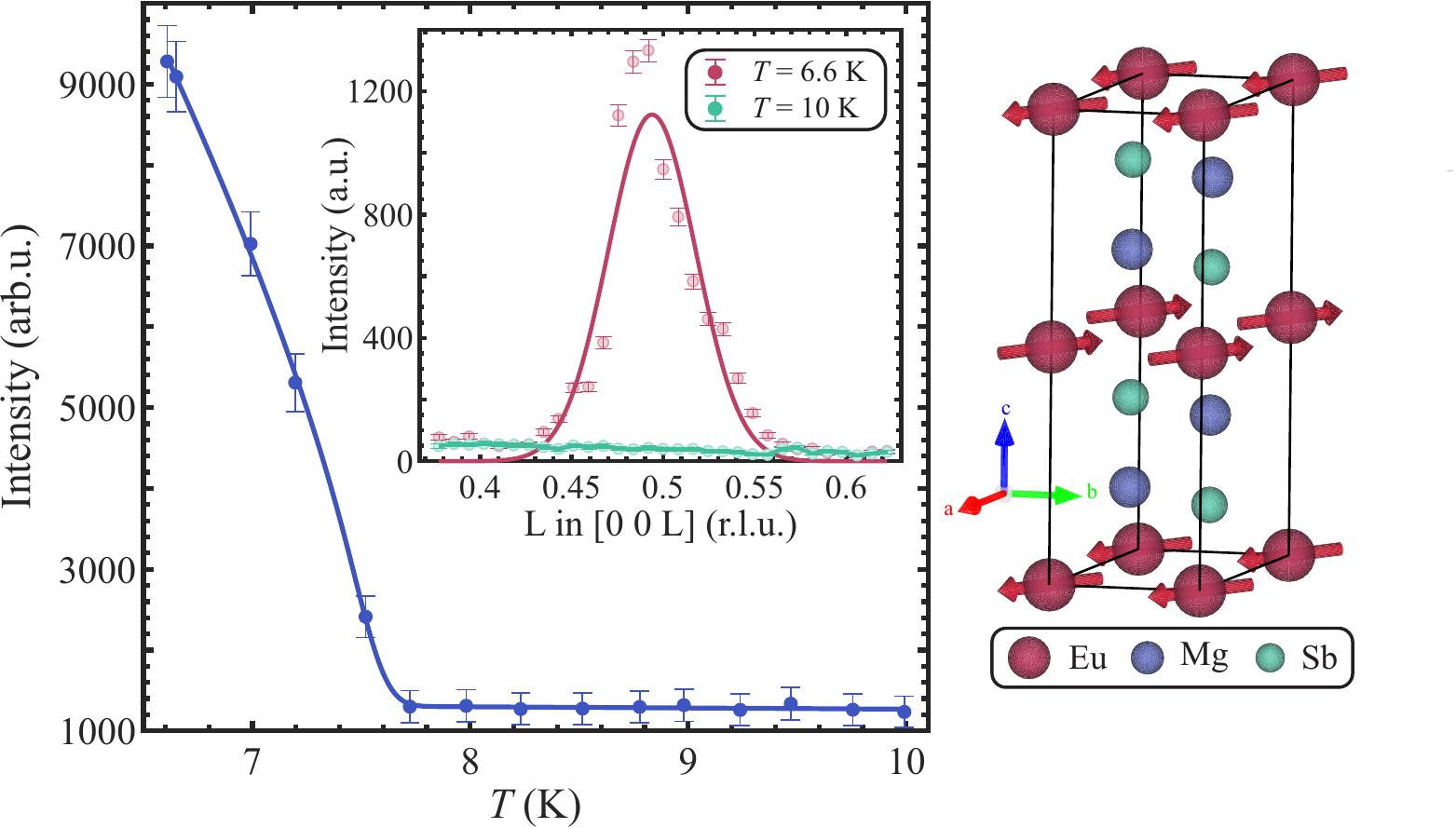}
\caption{Left panel: Integrated intensity as a function of temperature $T$ of the (0 0 $\frac{1}{2}$) magnetic Bragg reflection with the power-law fit~(\ref{Eq:power-law fit}) (solid blue line) yielding $T_{\rm N} = (7.7 \pm 0.4)$~K\@ and $\beta = 0.36 \pm 0.05$.  The error of $T_{\rm N}$ includes an estimated systematic error. Right panel: Chemical and A-type AFM spin structure of \ems. Our neutron-diffraction data are insensitive to the direction of the FM moment in the $ab$~plane although we show them to be pointing along the next-nearest-neighbor direction.}
\label{Fig:OP}
\end{figure}

The proposed A-type AFM structure is shown in the right panel of Fig.~\ref{Fig:OP}, where adjacent NN FM layers along the $c$~axis are rotated by 180$^\circ$ with respect to each other.   The direction of the FM moment within an Eu layer cannot be determined from neutron diffraction alone.  As noted for  isostructural EuMg$_2$Bi$_2$~\cite{Pakhira2021}, the only possible directions are towards NN (a,a1) or next nearest neighbor (NNN) (b,b1)  according to the Bilbao crystallographic server~\cite{Mato2015}.  Using published values for the structural parameters, we obtain good agreement with the intensities of the nuclear Bragg peaks, both above and below $T_{\rm N}$.  Based on this, we are able to confirm the A-type magnetic structure and obtain an estimate for the Eu ordered magnetic moment $g\langle S\rangle~\mu_{\rm B} = 4.0(5)~\mu_{\rm B}$ at $T = 6.6$ K by calculating the magnetic and chemical structure factors, where $S$ is the spin magnetic quantum number, $g$ is the spectroscopic-splitting factor, and $\mu_{\rm B}$ is the Bohr magneton.

As noted, our refinement of the magnetic structure yields an average magnetic moment at $T = 6.6$~K given by
\bea
\mu(6.6~{\rm K}) = \langle gS\rangle\,\mu_{\rm B} = (4.0\pm0.5)\,\mu_{\rm B}.
\eea
This value is smaller than the zero-temperature ordered moment \mbox{$\mu = gS\mu_{\rm B}=7\,\mu_{\rm B}$} expected from the electronic configuration of Eu$^{2+}$~\cite{Cable1977} with angular-momentum quantum number $L = 0$, spin quantum number $S=7/2$, and spectroscopic-splitting factor $g=2$ because  $\mu$  is not yet saturated to its full value at $T = 0$~K (see the left panel of Fig.~\ref{Fig:OP}).

The left panel of Fig.~\ref{Fig:OP} shows the integrated intensity of the \mbox{(0 0 $\frac{1}{2}$)} magnetic peak as a function of temperature where we used a simple power-law function
\bea
I_{\rm (0\,0\, 0.5)}(T) = C|1-T/T_{\rm N}|^{2\beta} \propto \mu^2
\label{Eq:power-law fit}
\eea
to fit the data (solid blue line with sharp transition).  The smooth line around $T_{\rm N}$ is obtained by the same power law but weighted by a Gaussian distribution of $T_{\rm N}$ (this form is sometimes used to account for crystal inhomogeneities) yielding $T_{\rm N} = (7.7 \pm 0.4$)~K and $\beta = 0.36 \pm 0.05$. The $T_{\rm N}$ is consistent within the error bars with $T_{\rm N}=8.0(2)$~K measured by  $\chi(T)$ and 8.0(1)~K measured by $C_{\rm p}(T)$ below. The power-law parameter~$\beta$ used here is phenomenological in order to determine $T_{\rm N}$.  Thus the value of $\beta$ is not reliable as a critical exponent. Importantly, the data and phenomenological fit in the left panel of Fig.~\ref{Fig:OP} show that the order parameter is still increasing strongly with decreasing~$T$ at $T = 6.6$~K and is therefore not close at that $T$ to its expected saturated value of $7\,\mu_{\rm B}$/Eu at $T=0$~K\@.

\subsection{\label{Sec:MagSus} Magnetic susceptibility}

\subsubsection{High-temperature regime}

\begin{figure}
\includegraphics [width=3in]{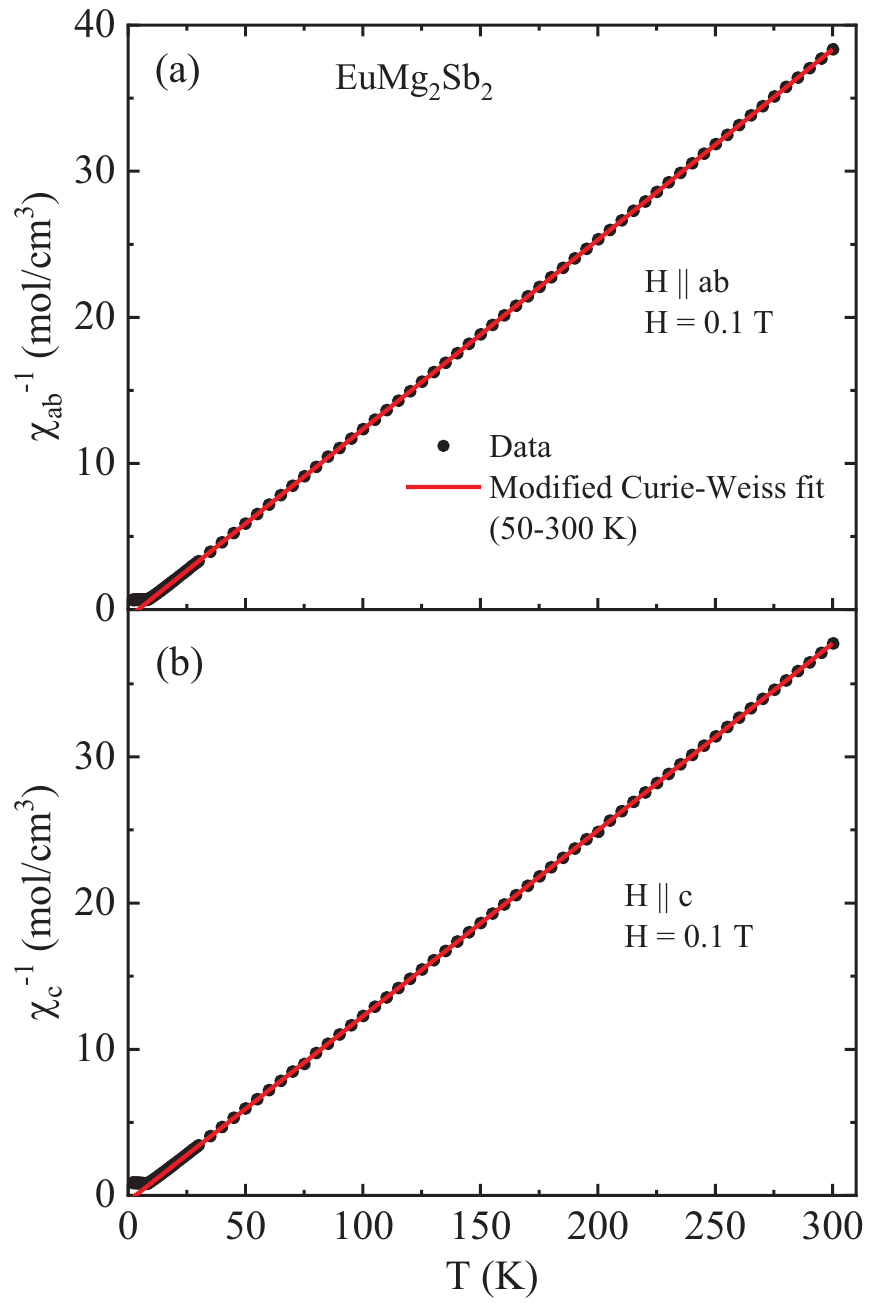}
\caption {Temperature dependence of the measured inverse magnetic susceptibilities (a) $\chi_{ab}^{-1}(T)$ and (b) $\chi_c^{-1}(T)$ (circles) along with the respective modified Curie-Weiss fits from 50 to 300~K (red straight lines).}
\label{Fig_Inverse_Chi}
\end{figure}

The inverse magnetic susceptibility data obtained in an applied field $H=0.1$~T are plotted for $H\parallel ab$ in Fig.~\ref{Fig_Inverse_Chi}(a) and for $H\parallel c$ in Fig.~\ref{Fig_Inverse_Chi}(b).  The data in the paramagnetic (PM) regime above  50~K for each field direction were fitted by the modified Curie-Weiss law
\bea
\chi_{\alpha}(T) =\chi_0 + \frac{C_{\alpha}}{T-\theta_{\rm p\alpha}} \quad (\alpha ~=~ab,~c),
\label{Eq.ModCurieWeiss}
\eea
where $\chi_0$ is an isotropic temperature-independent term, $\theta_{\rm p}$ is the Weiss temperature, and $C_\alpha$ is the Curie constant given by
\bse
\label{Eqs:Calpha_mueff}
\be
C_{\alpha}=\frac{N_{\rm A} {g_\alpha}^2S(S+1)\mu^2_{\rm B}}{3k_{\rm B}} = \frac{N_{\rm A}\mu^2_{\rm {eff, \alpha}}}{3k_{\rm B}},
\label{Eq.Cvalue1}
\ee
where the effective magnetic moment is given by
\be
\mu_{\rm {eff, \alpha}} = g_\alpha \sqrt{S(S+1)}\,\mu_{\rm B},
\label{Eq.mueff}
\ee
\ese
where $N_{\rm A}$ is Avogadro's number. The fits to the $h\parallel ab$ and $H\parallel c$ data by Eq.~(\ref{Eq.ModCurieWeiss}) are shown as red solid lines in Figs.~\ref{Fig_Inverse_Chi}(a) and~\ref{Fig_Inverse_Chi}(b), respectively, and the fitted parameters for each field direction are listed in Table~\ref{Tab.chidata}. The effective moment values obtained from $C_\alpha$ for both applied field directions are close to the value 7.94~$\mu_{\rm B}$/Eu expected from Eq.~(\ref{Eq.mueff}) for Eu$^{2+}$ spins with $S = 7/2$ and $g = 2$. The positive values of $\theta_{ab}$ and $\theta_{c}$ indicate dominant ferromagnetic (FM) interactions between the Eu spins, consistent with the A-type AFM structure obtained from the neutron-diffraction measurements where the in-plane ordered magnetic moments are ferromagnetically aligned and the nearest-neighbor moments along the $c$~axis are antiferromagnetically aligned.

\begin{table}
\caption{\label{Tab.chidata} Parameters obtained from fits of the data in Figs.~\ref{Fig_Inverse_Chi}(a) and~\ref{Fig_Inverse_Chi}(b) by Eqs.~(\ref{Eq.ModCurieWeiss}) and~(\ref{Eq.Cvalue1}).  Listed are the $T$--independent contribution to the susceptibility $\chi_0$, Curie constant per mol $C_\alpha$ in $\alpha = ab, c$ directions, effective moment per Eu $\mu{\rm_{eff}(\mu_B/Eu)} \approx \sqrt{8C}$ and Weiss temperature $\theta\rm_{p\alpha}$ obtained from the $\chi^{-1}(T)$ versus $T$ data for $H = 0.1$~T using Eq.~(\ref{Eq.ModCurieWeiss}).  The negative signs of the $\chi_0$ values indicate diamagnetic contributions, whereas the positive signs of the Weiss temperatures $\theta_{\rm p\alpha}$ indicate dominant ferromagnetic interactions.}
\begin{ruledtabular}
\begin{tabular}{ccccc}	
Field  & $\chi_0$ 				& $C_{\alpha}$ 		    &  $\mu_{\rm eff\alpha}$ 	& $\theta_{\rm p\alpha}$ \\
direction	& $\rm{\left(10^{-4}~\frac{cm^3}{mol}\right)}$	 & $\rm{\left(\frac{cm^3 K}{mol}\right)}$    & $\rm{\left(\frac{\mu_B}{mol}\right)}$& (K)  \\
\hline
H $\parallel ab$ 		& $-2.2(2)$		&  	7.77(1)	&	7.88(1)		& 4.32(6)   \\
H $\parallel c$ 	    		& $-2.0(5)$ 		&  	7.92(2)	&	7.95(1)	    	& 3.2(1)	\\
\end{tabular}
\end{ruledtabular}
\end{table}

Now we make a rough estimate of the nearest-neighbor exchange interactions $J_{ab}$ in the $ab$~plane and along the $c$~axis $J_c$, where a positive value is AFM and a negative value is FM\@.  Further-neighbor interactions are neglected.  According to molecular field theory (MFT)~\cite{Johnston2015}, for a lattice of spins that are identical and crystallographically equivalent, one has
\bse
\label{Eqs:ThetaTN}
\bea
\theta_{\rm p} &=& -\frac{S(S+1)}{3k_{\rm B}}\sum_jJ_{ij},\\
T_{\rm N} &=& -\frac{S(S+1)}{3k_{\rm B}}\sum_jJ_{ij}\cos(\phi_{ji}),
\eea
\ese
where the sums are over all neighbors~$j$ of a central spin~$i$, $J_{ij}$ is the exchange interaction between spins $i$ and $j$, and $\phi_{ji}$ is the angle between spins $j$ and $i$ in the magnetically-ordered state.  Here we only consider the nearest-neighbor (NN) spins to a central spin~$i$, of which there are 6 NN spins in the $ab$~plane with expected FM (negative) exchange interaction $J_{ab}$ and two NN spins along the $c$~axis with expected AFM (positive) exchange interactions.  Then Eqs.~(\ref{Eqs:ThetaTN}) become
\bse
\label{Eqs:FindThetTN}
\bea
\theta_{\rm p} &=& -\frac{S(S+1)}{3k_{\rm B}}(6J_{ab}+2J_c),\\
T_{\rm N} &=& -\frac{S(S+1)}{3k_{\rm B}}(6J_{ab}-2J_c).
\eea
\ese
Taking the average value $\theta_{\rm p,ave}= 3.9$~K from Table~\ref{Tab.chidata}  and $T_{\rm N}=8.0$~K, Eqs.~(\ref{Eqs:FindThetTN}) yield FM $J_{ab} = -0.016$~meV and AFM $J_c = 0.017$~meV, with the expected FM and AFM signs of $J_{ab}$ and $J_{c}$, respectively.

The relationship $\theta_{ab}>\theta_{c}$  in Table~\ref{Tab.chidata} and the preference for ordering of the Eu spins in the $ab$ plane as opposed to along the $c$ axis likely both arise at least in part from magnetic dipole interactions between the ordered Eu moment which strongly favor $ab$-plane moment alignment over \mbox{$c$-axis} alignment in the stacked triangular Eu lattice of \ems\ shown in the right panel of Fig.~\ref{Fig:OP} as follows.

In general, the eigenenergies $E_{i\alpha}$ of the magnetic dipole interaction tensor describing the interactions of magnetic dipole~$i$ with all other dipoles in an infinite crystal with a collinear magnetic structure is given by~\cite{Johnston2016}
\bse
\label{Eqs:Eialpha}
\bea
E_{i\alpha} = -\varepsilon\ \lambda_{{\bf k}\alpha},
\label{Eq:Ealpha}
\eea
where
\bea
\varepsilon = \frac{\mu^2}{2a^3},
\label{Eq:eps}
\eea
\ese
$\mu$ is the magnitude of the magnetic moment on each magnetic atom, $a$ is the crystallographic lattice constant in the $ab$~plane of the lattice of magnetic dipoles under consideration, and $\lambda_{{\bf k}\alpha}$ is the eigenvalue of the magnetic-dipole-interaction tensor for a particular magnetic propagation vector {\bf k} where $\alpha$ is the Cartesian principal-axis ordering direction of the collinear magnetic structure and hence the three principal axes are orthogonal to each other.  In general, the value of $\lambda_{{\bf k}\alpha}$ depends on the $c/a$ ratio of the particular magnetic structure under consideration except for cubic structures for which $c=a$.

\begin{figure}
\includegraphics [width=3.in]{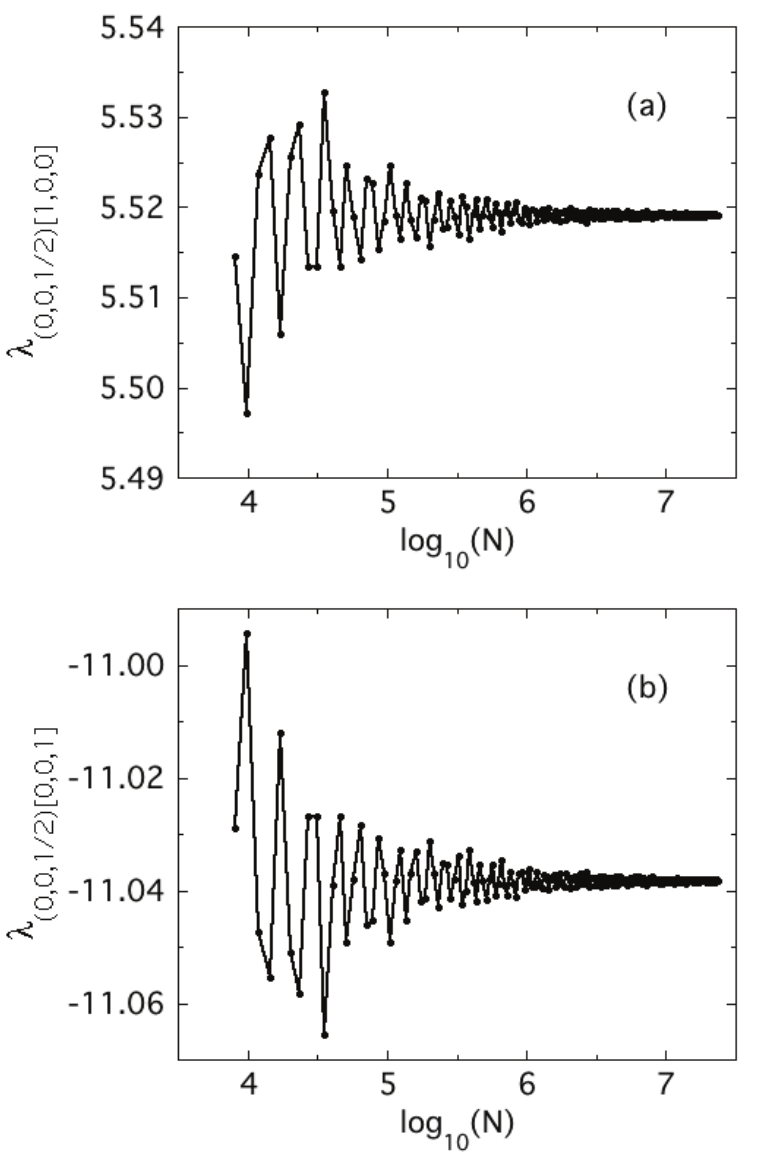}
\caption {Convergence of the magnetic dipole interaction tensor eigenvalues $\lambda_{k\alpha}$ for A-type AFM with A-type AFM propagation vector (0,0,1/2)~r.l.u.\ and moment alignment along (a)~the $a$~axis and (b)~the $c$~axis.}
\label{Fig_dipole_calcs}
\end{figure}

As discussed in Sec.~\ref{Sec:ExpDet}, the trigonal structure of \ems\ has lattice parameters $a=4.6531$\,\AA\ and \mbox{$c=7.6668$}\,\AA\ at $T=6.6$~K, yielding $c/a=1.6477$.  Using the Eu moment $\mu=7\,\mu_{\rm B}$,  Eq.~(\ref{Eq:eps}) gives
\bea
\varepsilon = 2.0916\times10^{-17}~{\rm erg}.
\eea
Following Ref.~\cite{Johnston2016}, for A-type ordering with AFM propagation vector {\bf k} = (0,0,1/2)~r.l.u.\ for the stacked triangular Eu spin lattice and after summing over a sphere of radius 200~$a$ containing $N_{\rm max} = 2.35\times 10^7$ spins, we find
\bse
\bea
\lambda_{(0,0,1/2)[1,0,0]} &=&  5.519\quad (a~{\rm axis~ordering)},\\
\lambda_{(0,0,1/2)[0,0,1]} &=&  -11.038\quad (c~{\rm axis~ordering)}.
\eea
\ese
These values are close to the values for a single FM triangular-lattice layer given by $\lambda_{[1,0,0],[0,1,0]}=5.517\,088$ and $\lambda_{[0,0,1]} = -11.034\,176$~\cite{Johnston2016}, as expected from the dependence of these values on the $c/a$ ratio as shown in the Appendix.  The differences here are
\bse
\bea
\lambda_{(0,0,1/2)[100]} - \lambda_{(0,0,1/2)[001]} &=& 16.557,\label{Eq:DeltaE}\\
\lambda_{(0,0,1/2)[100]} - \lambda_{(0,0,1/2)[010]} & \approx& -2.2\times10^{-12}.\label{Eq:abPlaneAnis}
\eea
\ese
The respective convergences of $\lambda_{(0,0,1/2)[100]}$ and $\lambda_{(0,0,1/2)[001]}$ versus log$_{10}$ of the number of spins~$N$ in the sphere of calculation are shown in Figs.~\ref{Fig_dipole_calcs}(a) and~\ref{Fig_dipole_calcs}(b), respectively, where the accuracy of convergence of \mbox{$\approx \pm 0.001$} for each $\lambda$ for the above value of $N_{\rm max}$ are seen.

According to Eqs.~(\ref{Eq:Ealpha}) and~(\ref{Eq:DeltaE}), $ab$-plane ordering is favored over $c$-axis ordering at $T=0$~K by an energy
\bea
\Delta E = 3.463\times10^{-16}~{\rm erg} = 0.2162~{\rm meV}.
\eea
This energy difference corresponds to a field difference $\Delta H = \Delta E/\mu = 5335$~Oe, where $\mu=7\,\mu_{\rm B}$/Eu.  This value is similar to the difference in the critical fields at \mbox{$T=1.8$~K} in Fig.~\ref{Fig_M-H_1p8K}(a) below, where \mbox{$H^{\rm c}_c~(H\parallel c)=3.4(1)$~T} and $H^{\rm c}_{\rm ab}~(H\parallel ab)=  2.6(1)$~T, yielding $H^{\rm c}_c-H^{\rm c}_{ab}=8$~kOe.  This indicates that the magnetic-dipole interaction is an important source of magnetic anisotropy in \ems.

On the other hand, the magnetic-dipole anisotropy within the $ab$~plane in Eq.~(\ref{Eq:abPlaneAnis}) is 13 orders of magnitude smaller than between the $ab$~plane and the $c$~axis in Eq.~(\ref{Eq:DeltaE}), where according to Eqs.~(\ref{Eq:Ealpha}) and (\ref{Eq:abPlaneAnis}), ordering perpendicular to the $a$~axis is slightly favored over ordering parallel to the $a$~axis.  It would be interesting to calculate the in-plane anisotropy using other methods.

\subsubsection{Low-temperature regime}

\begin{figure}
\includegraphics [width=3in]{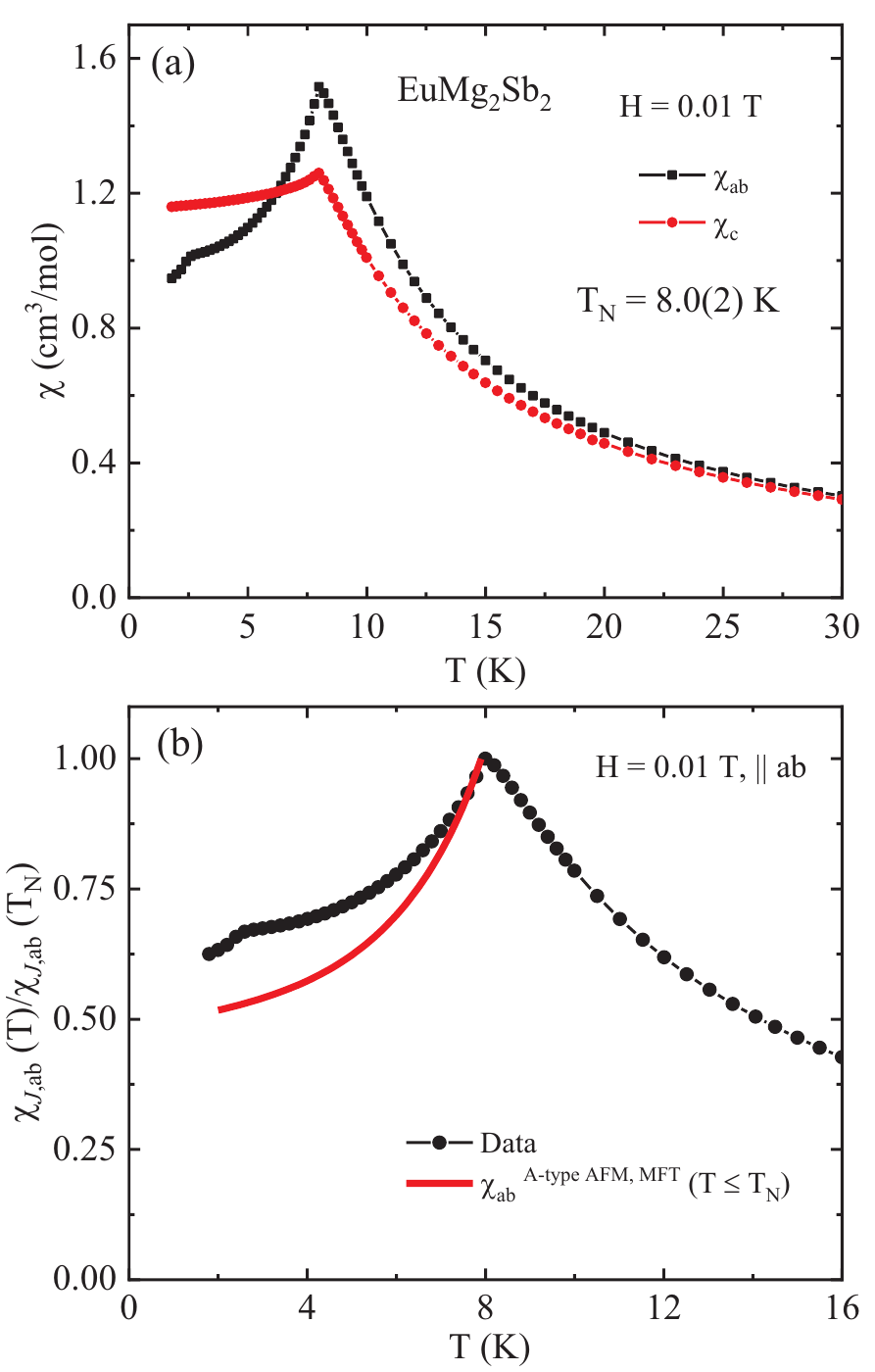}
\caption {(a) Magnetic susceptibility $\chi = M/H$ of \ems\ as a function of $T$ below 30~K for $H \parallel ab$ and $H \parallel c$ measured in $H = 0.01$~T\@.  The $\chi_{ab}(T)$ and $\chi_{c}(T)$ are seen to be anisotropic for $T \lesssim 30$~K, which is greater than $T_{\rm N} = 8.0(2)$~K\@. (b)~The ratio $\chi_{ab}(T)/\chi_{ab}(T_{\rm N})$ for $H = 0.01$~T (black symbols). The red solid line is the $\chi_{Jab}(T)/\chi_{Jab}(T_{\rm N})$ calculated for A-type AFM order for $T < T_{\rm N}$ according to MFT~\cite{Johnston2012, Johnston2015}. }
\label{Fig_M-T}
\end{figure}

\begin{figure}
\includegraphics [width=3in]{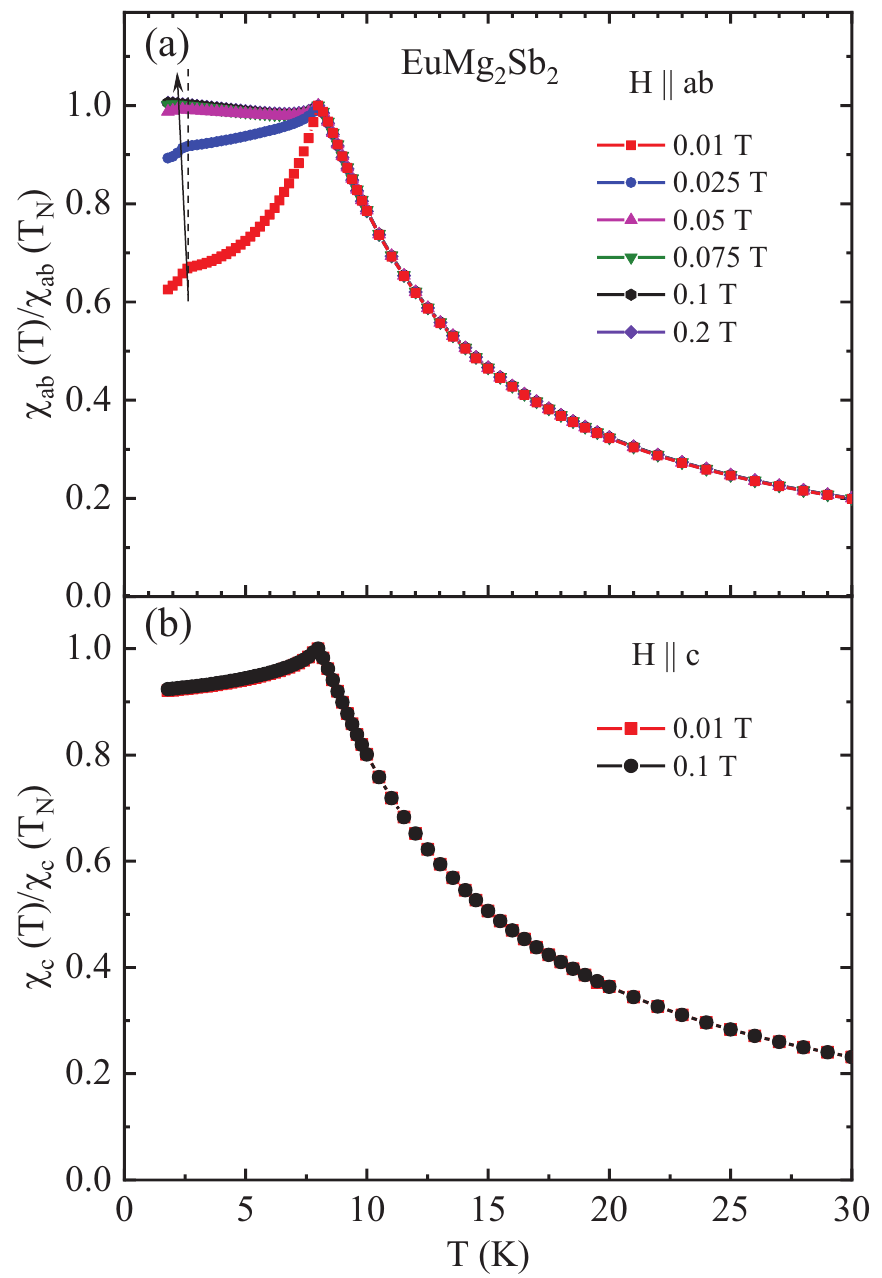}
\caption {Temperature dependence of $\chi_{ab}$ and $\chi_{c}$ of \ems\ normalized to that at $T = T_{\rm N}$ for different applied magnetic fields in (a) and (b), respectively.  In~(a), increasing fields are indicated by an arrow.   $\chi_{ab}$ for $T < T_{\rm N}$ is seen to strongly increase with increasing~$H$ before becoming independent of $T$ and $H$ for $H \gtrsim 0.075$~T, whereas $\chi_{c}$ is  independent of $H$ and $T$ below $T_{\rm N}$ for fields of 0.01 and 0.1~T\@.  The kink at $T = 3$~K in the $\chi_{ab}(T)$ data of unknown origin measured at $H = 0.01$~T shifts to lower $T$ with increasing $H$\@. The dashed vertical line at constant $T$ is included in (a) to more clearly show the small $H$ dependence of the transition at 3~K.}
\label{Fig_M-T_diff_fields}
\end{figure}

Figure~\ref{Fig_M-T}(a) shows $\chi_{ab}(T)$ and $\chi_c(T)$ of \ems\ measured in $H = 0.01$~T for $T\leq 30$~K\@.  Although $\chi$ is isotropic at $T> 30$~K, anisotropy is seen below 30~K which increases with decreasing $T$\@.  The compound exhibits a clear AFM transition on cooling below $T_{\rm N} = 8.0(2)$~K, where sharp peaks are observed at this temperature in both $\chi_{ab}(T)$ and $\chi_{c}(T)$. Below $T_{\rm N}$, $\chi_{ab}(T)$ decreases with decreasing~$T$, whereas $\chi_{c}(T)$ is almost independent of $T$ indicating moment alignment in the $ab$~plane, consistent with the A-type AFM structure obtained from the above neutron-diffraction measurements in which the moments are aligned ferromagnetically in the $ab$~plane.  A magnetic transition of unknown type also appears as a cusp at 3~K in the $\chi{ab}(T)$ data, but there is no evidence of this transition in the $\chi_c(T)$ data.  The entropy change associated with this phase transition is evidently quite small, because our $C_{\rm p}(T)$ data in Sec.~\ref{Sec:Heatcap} below show no clear evidence for a phase transition at this temperature.

The $\chi_{ab}(T)$ normalized by the value of $\chi_{ab}(T_{\rm N})$ is shown in Fig.~\ref{Fig_M-T}(b), together with expectation for A-type AFM order (red curve) to be discussed below.  The magnetic-field dependence of $\chi(T)/\chi(T_{\rm N})$ below 30~K  is shown in Figs.~\ref{Fig_M-T_diff_fields}(a) and~\ref{Fig_M-T_diff_fields}(b) for $H \parallel ab$ and $H \parallel c$, respectively.   Figure~\ref{Fig_M-T_diff_fields}(a) shows that the transition seen in $\chi_{ab}(T)$ at $T=3$~K is suppressed by an in-plane magnetic field.  Figure~\ref{Fig_M-T_diff_fields}(a) also shows that $\chi_{ab}(T)$ for $T < T_{\rm N}$ is strongly enhanced even at low fields and becomes independent of~$T$ for $H \gtrsim 0.075$~T\@.  We propose below that this behavior results from a field-induced magnetic-moment reorientation within the $ab$~plane.  In contrast, $\chi_{c}(T)$ does not change in $H=0.1$~T applied along the $c$~axis compared to the behavior in $H= 0.01$~T\@.

Our zero-field neutron-diffraction results indicate that the AFM structure from $T_{\rm N}$ to 6.6~K is A-type, where the magnetic moments are aligned ferromagnetically in the $ab$~plane and the moments in adjacent Eu planes along the $c$~axis are aligned antiferromagnetically. An A-type antiferromagnet with $ab$-plane moment alignment and turn angle $kd=180^\circ$ between adjacent layers is equivalent to a $c$-axis helix with a turn angle $kd\to 180^\circ$.

Here, we calculate $\chi_{ab}(T\leq T_{\rm N})$ for such an A-type AFM structure using MFT~\cite{Johnston2012, Johnston2015}.  To use this theory, the direction-averaged $\chi_{Jab}(T)$ for $T\geq T_{\rm N}$ is defined as
\bea
\chi_{Jab}(T) = \frac{2}{3}\chi_{ab}(T)+ \frac{1}{3}\chi_c(T),
\eea
Then the $\chi_{ab}(T\leq T_{\rm N})$ data are shifted vertically to agree at $T_{\rm N}$ with $\chi_{Jab}(T_{\rm N})$.  We designate the resulting data at $T\leq T_{\rm N}$ also as $\chi_{Jab}(T)$ and then form the ratio $\chi_{Jab}(T)/\chi_{Jab}(T_{\rm N})$ as plotted versus $T$ in Fig.~\ref{Fig_M-T}(b).

With the definition $f=\theta_{\rm p\,ave}/T_{\rm N} = 0.49$ from Table~\ref{Tab.chidata}, we have~\cite{Johnston2012, Johnston2015}
\bse
\label{Eqs:Chixy}
\be
\frac{\chi_{Jab}(T \leq T_{\rm N})}{\chi_J(T_{\rm N})}=  \frac{(1+\tau^*+2f+4B^*)(1-f)/2}{(\tau^*+B^*)(1+B^*)-(f+B^*)^2},
\label{eq:Chi_plane}
\ee
where
\be
B^*=  2(1-f)\cos(kd)\,[1+\cos(kd)] - f,
\label{eq:Bstar}
\ee
\be
t =\frac{T}{T_{\rm N}},\quad \tau^*(t) = \frac{(S+1)t}{3B'_S(y_0)}, \quad y_0 = \frac{3\bar{\mu}_0}{(S+1)t},
\ee
the ordered moment versus $T$ in $H=0$ is denoted by $\mu_0$, the reduced ordered moment $\bar{\mu}_0 = \mu_0/\mu_{\rm sat}$ is determined by numerically solving the self-consistency equation
\be
\bar{\mu}_0 = B_S(y_0),
\label{Eq:barmuSoln}
\ee
$B'_S(y_0) = [dB_S(y)/dy]|_{y=y_0}$, and the Brillouin function $B_S(y)$ is
\be
B_S(y)= \frac{1}{2 S}\left\{(2S+1){\rm coth}\left[(2S+1)\frac{y}{2}\right]-{\rm coth}\left(\frac{y}{2}\right)\right\}.
\ee
\ese
At $T=0$, one obtains~\cite{Johnston2012, Johnston2015}
\be
\label{Eq:kd}
\frac{\chi_{Jab}(T=0)}{\chi_{Jab}(T_{\rm N})}=\frac{1}{2[1+2~{\rm cos}(kd)+2~{\rm cos}^2(kd)]}.
\ee
Then taking the turn angle between moments in adjacent layers of Eu spins as $kd\to180^\circ$ for the \mbox{A-type} AFM structure in \ems\ discussed above with the moments aligned in the $ab$~plane gives \mbox{$\chi_{Jab}(T=0)/\chi_{Jab}(T_{\rm N}) \to 1/2$}, as typically observed.

The $\chi_{Jab}(T)/\chi_{Jab}(T_{\rm N})$ calculated for \ems\ is shown as the red curve in Fig.~\ref{Fig_M-T}(b). However, as seen from the figure, the experimental data appreciably differ from the predicted behavior for A-type antiferromagnetism, evidently due to presence of the additional transition at $T = 3$~K and the associated influence on $\chi_{ab}(T)$ at $T<T_{\rm N}$. We could not probe the magnetic structure below 3~K using neutron diffraction because the cryostat used has a low-$T$ limit of 6.6~K\@.  Future neutron-diffraction experiments at lower temperatures would be of great interest.

\subsection{\label{Sec:IsoMag} Isothermal magnetization versus field}

\subsubsection{Overview}

\begin{figure}[h!]
\includegraphics [width=3.3in]{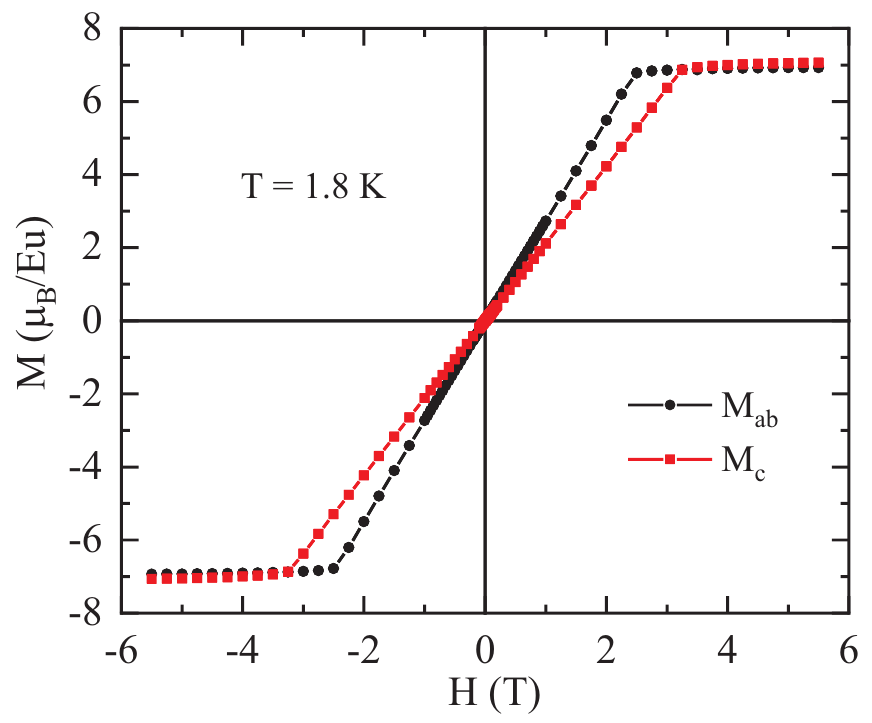}
\caption {Magnetic hysteresis curves measured at $T = 1.8$~K for $H \parallel ab$ ($M_{ab}$) and $H \parallel c$ ($M_{c}$).}
\label{Fig_M-H_1p8K}
\end{figure}

\begin{figure}
\includegraphics [width=3in]{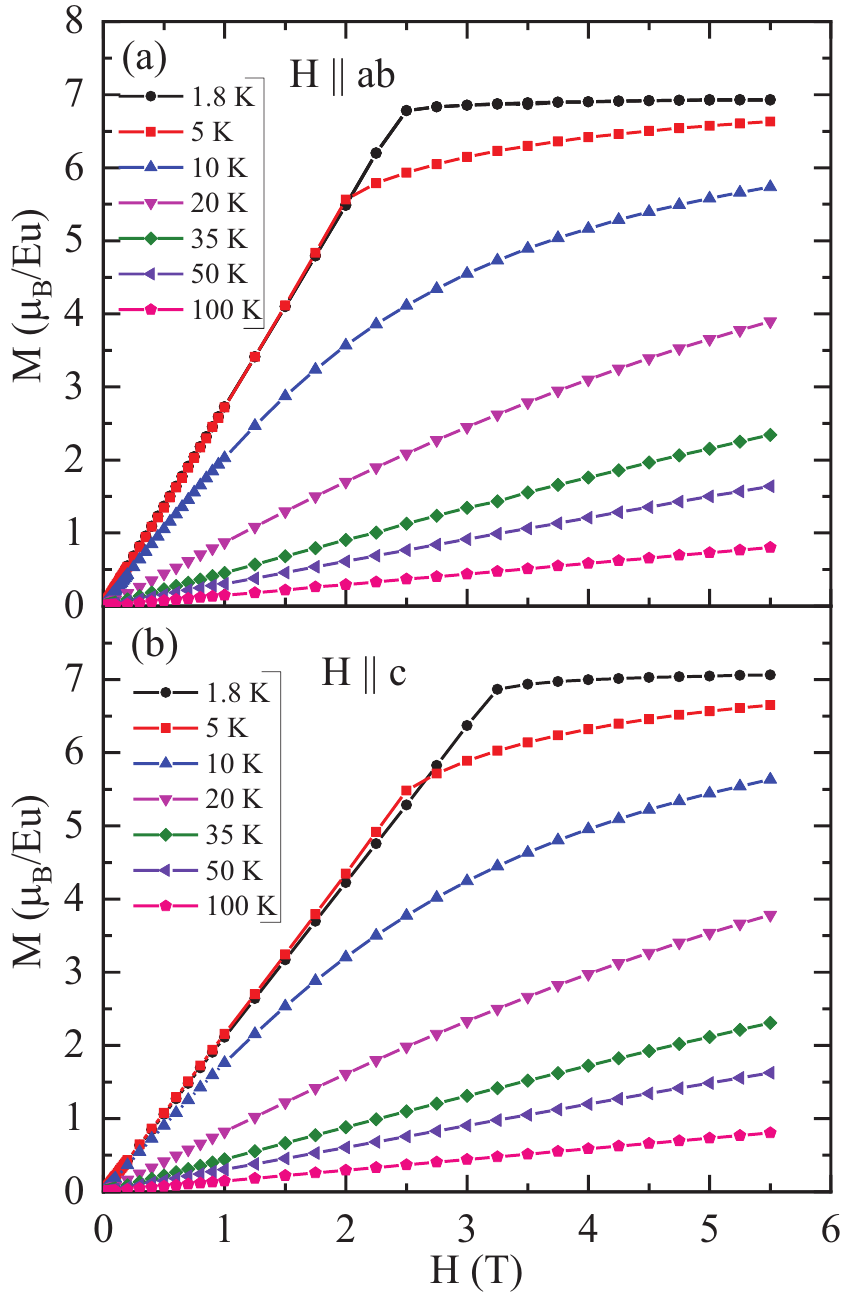}
\caption {Field dependence of magnetic isotherms measured at different temperatures when the applied field is (a) in the $ab$ plane and (b) along the $c$~axis.}
\label{Fig_M-H_diff_temp}
\end{figure}

\begin{figure}[h!]
\includegraphics [width=3.3in]{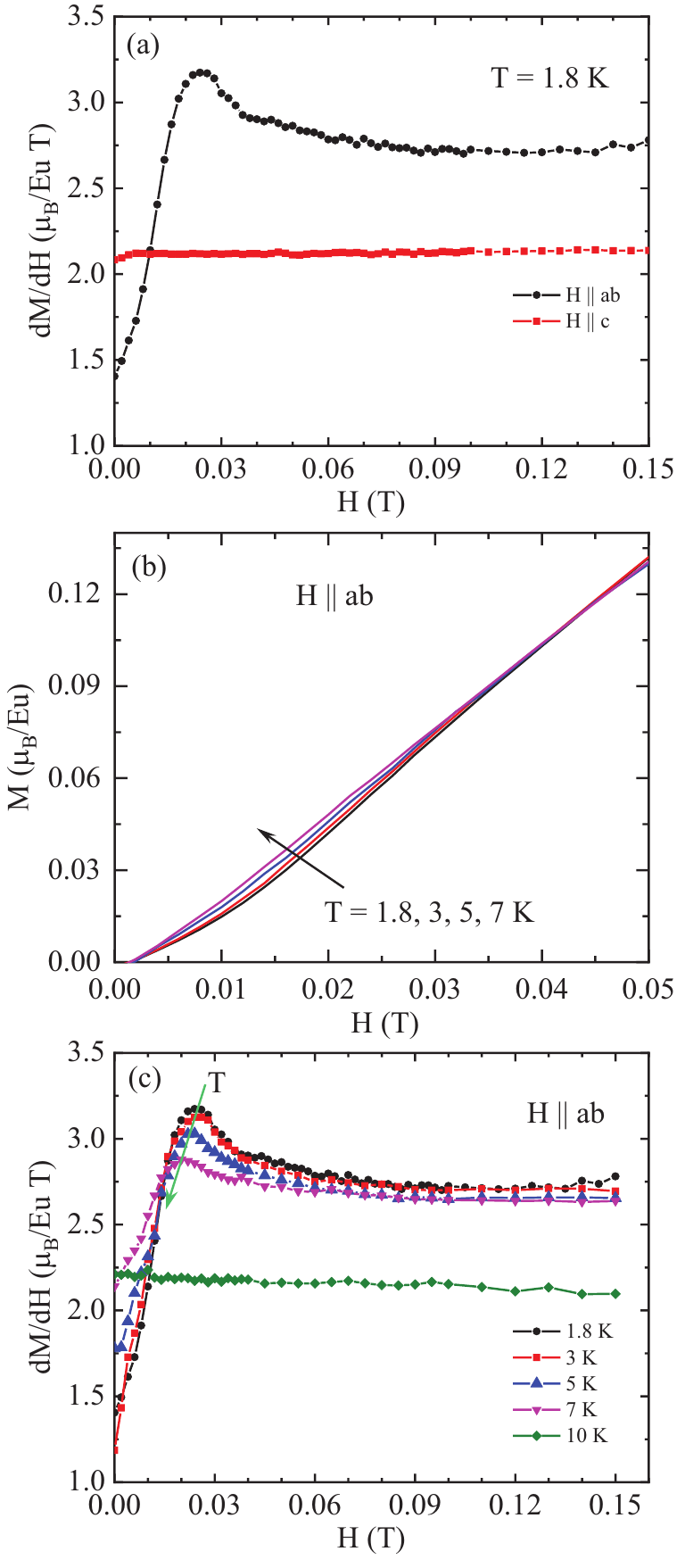}
\caption {(a)~Low-field $dM/dH$ vs. $H$ from Fig.~\ref{Fig_M-H_1p8K}. The data for $H \parallel ab$ exhibit a distinct nonlinearity. (b)~Temperature dependence of $M_{ab}(H)$ emphasizing the low-field region; the corresponding $dM/dH$ versus~$H$ is presented in (c).}
\label{Fig_M-H_low_field}
\end{figure}

In order to provide further insight into the field-induced evolution of the magnetic properties, isothermal magnetization versus field $M(H)$ measurements were carried out on \ems\ crystals. \mbox{Figure~\ref{Fig_M-H_1p8K}} shows  $M(H)$ hysteresis curves for \mbox{$-5.5~{\rm T} \leq H \leq 5.5$~T}  measured at \mbox{$T = 1.8$~K} for both $H \parallel ab$ ($M_{ab}$) and $H \parallel c$ ($M_{c}$). Over this field range, both $M_{ab}$ and $M_{c}$ appear to increase linearly with increasing field and saturate above the critical fields $H^{\rm c}_{ab} = 2.6(1)$~T and $H^{\rm c}_c = 3.4(1)$~T\@. The saturation moment $\mu_{\rm sat} =  7.0(5)\,\mu_{\rm B}$/Eu at $T = 1.8$~K is observed for both field directions, which agree within the errors with the theoretical value $\mu = gS\mu_{\rm B} = 7~\mu_{\rm B}$/Eu expected for Eu$^{2+}$ with $g = 2$ and $S = 7/2$.

Figure~\ref{Fig_M-H_diff_temp} shows the temperature evolution of the $M(H)$ isotherms from 1.8 to 100~K\@.  The data at 5~K show a decrease in the critical fields to $H^{\rm c}_{ab} = 2.0$~T and $H^{\rm c}_c = 2.5$~T\@.  The data at higher temperatures are in the paramagnetic regime where the $M(H)$ data become linear.

\subsubsection{Magnetic moment reorientation in small $ab$-plane magnetic fields at low temperatures}

The derivative $dM/dH$ versus $H$ at $T=1.8$~K is plotted versus~$H$ in Fig.~\ref{Fig_M-H_low_field}(a) for both $H\parallel ab$ and $H\parallel c$.  The data show that $M_{ab}(H)$  is nonlinear in the low-field region $H\lesssim 0.1$~T, whereas $M_{c}(H)$ is linear. The temperature dependence of the nonlinear $M_{ab}(H)$ behavior is shown in Fig.~\ref{Fig_M-H_low_field}(b) and the corresponding field derivatives in Fig.~\ref{Fig_M-H_low_field}(c). It is evident that the nonlinearity in $M_{ab}(H)$ persists up to $T_{\rm N} = 8$~K with a maximum slope at $H \sim 0.025$~T for $T = 1.8$~K, where a peak in $dM_{ab}/dH$ is observed. The peak shifts to lower fields with increasing $T$\@. This behavior is similar to our earlier observations for two other trigonal Eu-based compounds \emb\ and \esa~\cite{Pakhira2020, Pakhira2021a}, where we argued that the nonlinearity results from magnetic-field-induced ordered-moment reorientation in the three trigonal AFM domains associated with a very weak in-plane magnetic anisotropy. This scenario is plausible here as well. The antiparallel AFM spins in each  of the three domains present at zero-field rotate to become perpendicular to {\bf H} at $H \sim 0.06$~T, apart from a small canting towards {\bf H} to produce the observed magnetization.

\subsection{\label{Sec:Heatcap} Heat capacity}

\begin{figure}
\includegraphics [width=3.3in]{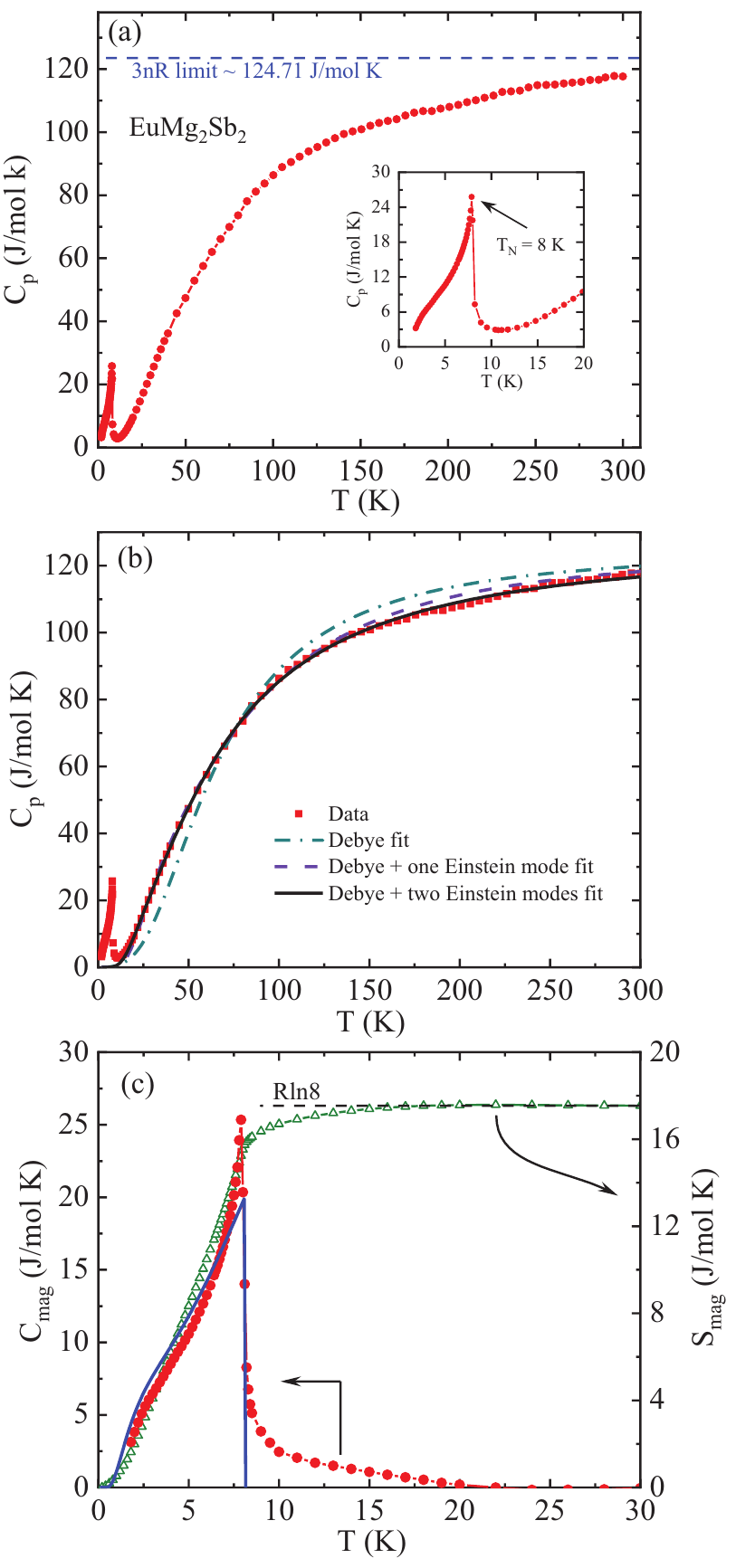}
\caption {(a) Temperature $T$ dependence of the zero-field heat capacity $C_{\rm p}$ of an \ems\ single crystal. The horizontal blue dashed line is the high-$T$ Dulong-Petit limit.  Inset: Expanded plot showing the $\lambda$-type peak at $T_{\rm N}=8.0(1)$~K\@. (b)~$C_{\rm p}(T)$ fitted above $T=50$~K by the Debye model with $\gamma = 0$ [blue dash-dotted line, Eq.~(\ref{Eq:Debye_Fit})] and with one [dashed purple line, Eqs.~(\ref{Eq:Debye and Einstein})] or two [solid black line, Eq.~(\ref{Eq:Debye and TwoEinstein})] Einstein contributions. (c)~Magnetic contribution $C_{\rm mag}(T)$ to $C_{\rm p}(T)$ (filled red circles, left ordinate) obtained by subtracting the fitted $C_{\rm p}(T)$ [Eq.~(\ref{Eq:Debye and TwoEinstein})] from the experimental data.  Also shown are the MFT prediction for $C_{\rm mag}(T)$ [Eq.~(\ref{Eq:deltaCp_T_MFT}), solid blue line] and the magnetic entropy $S_{\rm mag}(T)$ calculated from $C_{\rm mag}(T)$ using Eq.~(\ref{Smag}) (green diamonds, right ordinate). }
\label{Fig_Heat_cap_zero_field}
\end{figure}

The temperature dependence of the zero-field heat capacity $C_{\rm p}$ of \ems\ is shown in Fig.~\ref{Fig_Heat_cap_zero_field}(a). A sharp $\lambda$-type peak is observed at $T_{\rm N} = 8.0(1)$~K. The $C_{\rm p}$ value of \ems\ at $T = 300$~K is $\approx$ 117.5 J/mol\,K, which is smaller than the expected classical Dulong-Petit high-$T$ limit $3nR = 124.71$~J/mol K for the compound shown as the horizontal dashed line in Fig.~\ref{Fig_Heat_cap_zero_field}(a), where $n=5$ is the number of atoms per formula unit and $R$ is the molar gas constant.

We first fitted the $C_{\rm p}(T)$ data for $T= 50$--300~K by the Debye model according to the general expression
\bea
\label{Eq:Debye_Fit}
C_{\rm p}(T) &=& \gamma T+ n~ C_{\rm V\,Debye}(T),\label{Eq:Debye_Fit} \\*
C_{\rm V}(T) &=& 9R \left(\frac{T}{\Theta_{\rm D}}\right)^3\int_{0}^{\Theta_{\rm D}/T}\frac{x^4e^x}{(e^x-1)^2} dx,\nonumber
\eea
where $\gamma$ is the Sommerfeld electronic specific heat coefficient and $\Theta_{\rm D}$ is the Debye temperature. The lower limit of 50~K for the fit was chosen to avoid any contribution from short-range magnetic ordering of the Eu spins above $T_{\rm N}$.  We used $\gamma = 0$ since \ems\ is a semiconductor.  The fitted Debye temperature is $\Theta_{\rm D} = $270(3)~K\@.  As shown in Fig.~\ref{Fig_Heat_cap_zero_field}(b), the $C_{\rm p}(T)$ is not fitted well using the Debye model.  The failure of the Debye model to fit the lattice heat capacity suggests the presence of one or more optic-phonon modes that would give Einstein contributions to the lattice heat capacity.

We therefore next fitted the $C_{\rm p}(T)$ data from 50 to 300~K with a combination of the Debye model and an Einstein contribution associated with a single optic-phonon mode according to
\bse
\bea
\label{Eq:Debye and Einstein}
C_{\rm p}(T) &=& (1 - \alpha)C_{\rm V\,Debye}(T) + \alpha C_{\rm V\,Einstein}(T),~~~
\eea
where
\bea
\label{Eq:Einstein}
C_{\rm V\,Einstein}(T) &=& 3R \left(\frac{\Theta_{\rm E}}{T}\right)^2\frac{e^{\Theta_{\rm E}/T}}{(e^{\Theta_{\rm E}/T} - 1)^2},~~~
\eea
\ese
$\Theta_{\rm E}$ is the Einstein temperature, and the parameter $\alpha$ determines the relative contributions of the Debye and Einstein components to the lattice heat capacity.  The fitted parameters were $\Theta_{\rm D} = 415(12)$~K, $\Theta_{\rm E} = 111(3)$~K, and $\alpha = 0.48(2)$. We found that although the fit in the low-$T$ region improved significantly, there is still a significant discrepancy between the theory and experimental data for $T \geq 150$~K as shown in Fig.~\ref{Fig_Heat_cap_zero_field}(b).

We next considered a model containing a combination of the  Debye model and two Einstein modes using the relation
\begin{eqnarray}
&C_{\rm p}(T)& = (1 - \alpha_1 - \alpha_2)C_{\rm V\,Debye}(\Theta_{\rm D}, T) + \label{Eq:Debye and TwoEinstein}\\
&& \alpha_1 C_{\rm V\,Einstein}(\Theta_{\rm E_1}, T) + \alpha_2 C_{\rm V\,Einstein}(\Theta_{\rm E_2}, T).\nonumber
\end{eqnarray}
The parameters obtained from a fit of the $C_{\rm p}(T)$ data from 50 to 300~K are $\Theta_{\rm D} = 309(18)$~K, $\Theta_{\rm E_1} = 94(5)$~K with $\alpha_1$ = 0.30(4), and $\Theta_{\rm E_2} = 749(80)$~K with $\alpha_2$ = 0.08(2).  As seen in Fig.~\ref{Fig_Heat_cap_zero_field}(b) (black curve), this is clearly the best fit to the data over the temperature range 50--300~K\@.

The magnetic contribution to the heat capacity $C_{\rm {mag}}(T)$ was obtained by subtracting the lattice contribution based on the model in Eq.~(\ref{Eq:Debye and TwoEinstein}) from the measured $C_{\rm p}(T)$ data over the temperature range 1.8 to 30~K\@.  The result is shown in Fig.~\ref{Fig_Heat_cap_zero_field}(c). According to MFT~\cite{Johnston2015}, $C_{\rm {mag}}(T)$ is given by
\bea
\label{Eq:deltaCp_T_MFT}
C_{\rm {mag}}(t) = R\frac{3S\overline{\mu}_0^2(t)}{(S + 1)t[\frac{(S + 1)t}{3B^{\prime}_S(t)} - 1]},
\eea
where the symbols are defined in Eqs.~(\ref{Eqs:Chixy}). The solid blue curve in Fig.~\ref{Fig_Heat_cap_zero_field}(c) depicts $C_{\rm {mag}}(T)$ calculated using Eq.~(\ref{Eq:deltaCp_T_MFT}) with $T_{\rm N}$ = 8~K and $S = 7/2$, where the agreement between theory and experiment is seen to be quite good, although the $\lambda$ shape of the measured heat capacity and the short-range ordering that is clearly observed in the data above $T_{\rm N}$ are, of course, not reproduced by MFT\@.

The magnetic entropy $S_{\rm {mag}}(T)$ was calculated according to
\bea
\label{Smag}
S_{\rm mag}(T) = \int_{0}^{T}\frac{C_{\rm {mag}}(T)}{T} dT.
\eea
The temperature dependence of $S_{\rm {mag}}$ is shown in Fig.~\ref{Fig_Heat_cap_zero_field}(c) (green triangles, right ordinate). $S_{\rm {mag}}(T)$ reaches the expected high-$T$ limit $S_{\rm mag} = R{\rm ln}(2S + 1) = 17.29$ J/mol~K for $S=7/2$ at $T \gtrsim 20$~K rather than at $T_{\rm N}$.  The short-range magnetic ordering above $T_{\rm N}$ noted above is responsible for this difference,  as reported previously for the similar Eu-based compounds EuMg$_2$Bi$_2$, EuSn$_2$As$_2$, EuCo$_2$P$_2$, and EuCo$_{2-y}$As$_2$~\cite{Pakhira2020, Pakhira2021a, Sangeetha2016, Sangeetha2018}.

\section{\label{Sec:Conclu} Summary}

We have grown high-quality single crystals of the layered compound \ems\ and have studied its crystallographic, magnetic, electronic-transport, and thermal properties. The compound crystallizes in the trigonal \cas-type crystal structure where the Eu atoms form a simple-triangular lattice in the $ab$~plane that is stacked along the $c$~axis.

The temperature dependence of zero-field electrical resistivity $\rho(T)$ indicates that \ems\ is a  narrow-gap semiconductor with an intrinsic energy gap \mbox{$E_g=0.37$~eV}\@. This semiconducting state is also evident from the ARPES measurements. Although the similar  isostructural compound \emb\ was recently reported to have a semimetallic electronic ground state, the formation of a narrow-gap semiconducting state in \ems, where Sb has a smaller spin-orbit coupling (SOC) compared to Bi, suggests an important role of SOC coupling in tuning the electronic states of these Zintl-phase compounds.

The magnetic susceptibility $\chi(T)$ reveals a paramagnetic to AFM transition below the N\'{e}el temperature $T_{\rm N} = 8.0(2)$~K\@. An additional sharp cusp is observed in the in-plane susceptibility $\chi_{ab}(T)$ at $T = 3.0$~K which shifts to lower temperature with increasing $H$, whereas no such feature is observed in the $\chi_{c}(T)$ data. In addition, no anomaly in the heat capacity is observed at this temperature.  The nature of this 3~K transition remains to be identified.

Our zero-field neutron-diffraction measurements in the temperature range 6--10~K showed that the AFM structure below $T_{\rm N}$ is A-type. In this magnetic structure, the Eu$^{2+}$ spins~7/2 within an $ab$ plane are aligned ferromagnetically within the plane with the Eu spins in adjacent Eu planes along the $c$~axis aligned antiferromagnetically.

The magnetization in fields both in the $ab$~plane and along the $c$~axis saturates to a value of $7.0(5) \mu_{\rm B}$/Eu at $T = 1.8$~K, consistent with expectation for $S= 7/2$ with $g=2$, above the critical fields $H^{\rm c}_{ab} = 2.6(1)$~T and $H^{\rm c}_{c} = 3.4(1)$ T at $T=1.8$~K\@.

A sharp $\lambda$-type peak is observed at $T_{\rm N}$ in the zero-field heat capacity $C_{\rm p}(T)$ data, consistent with the expected second-order nature of the long-range AFM transition. The release of the full magnetic entropy at a temperature of 30~K that is higher than $T_{\rm N}$ signifies the presence of significant short-range magnetic correlations above $T_{\rm N}$.

Magnetic field-dependent $\chi(T)$ and isothermal magnetization $M(H)$ data suggest that the A-type AFM ground state consists of threefold trigonal AFM domains, each containing antiferromagnetically-aligned magnetic moments in zero field, but where the moments reorient under the influence of a weak external $ab$-plane magnetic field to become nearly perpendicular to $H$ for $H \sim 0.06$~T, apart from the weak canting of the moments towards the field which gives rise to the observed magnetization. Such a weak field-induced spin reorientation indicates the presence of a very weak in-plane anisotropy.  The in-plane anisotropy arising from the magnetic-dipole interaction is indeed found to be extremely small.  On the other hand, the magnetic dipole interaction is found to be responsible for much of the difference between the critical magnetic fields measured in the $ab$~plane and along the $c$~axis.

\newpage

\appendix*

\section{Dependence of the Magnetic-Dipole Eigenvalues on the $c/a$ Ratio of the Stacked-Triangular Lattice with A-type Antiferromagnetic Order}

\begin{figure}[h]
\includegraphics [width=3.3in]{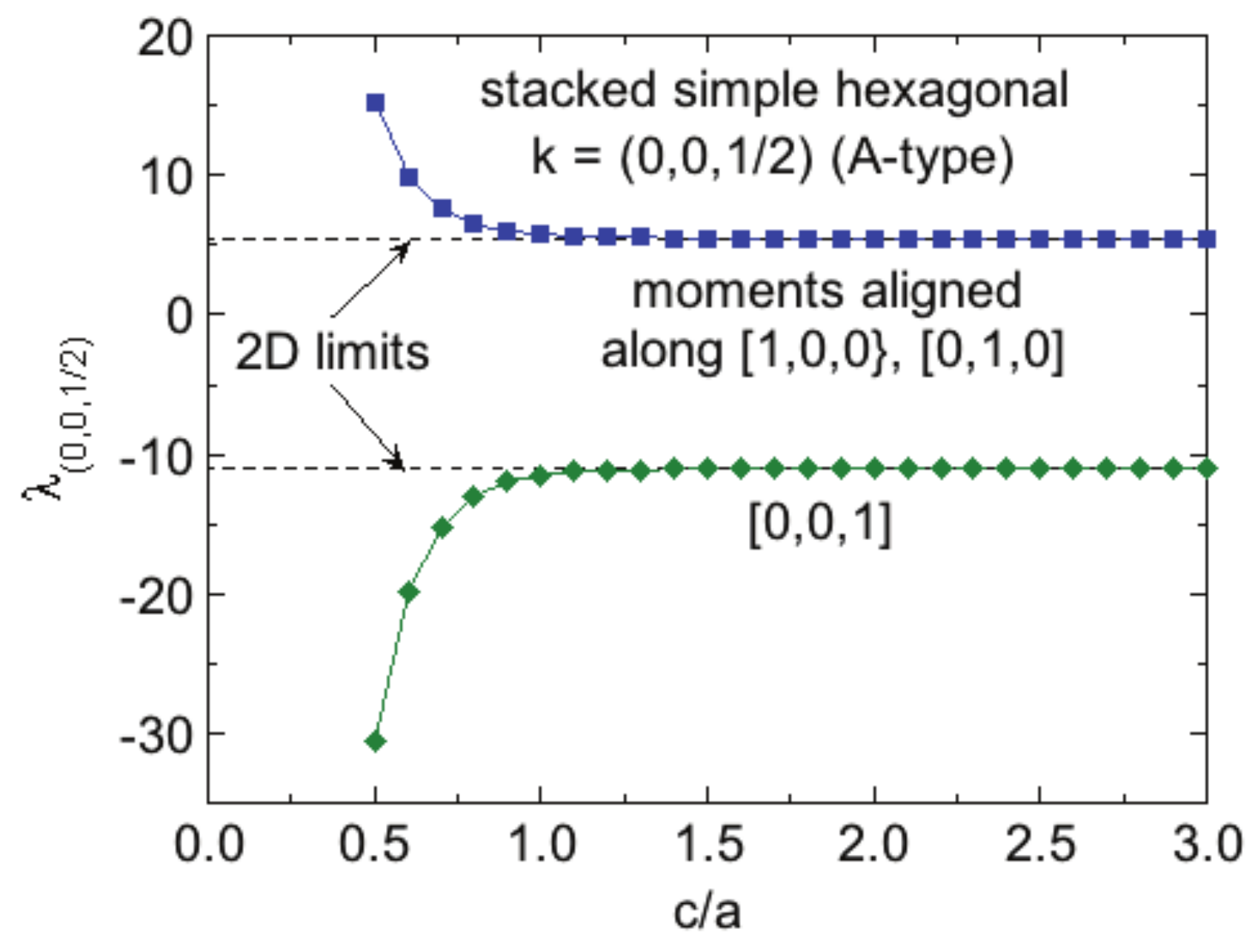}
\caption {Magnetic dipole eigenvalues $\lambda(0,0,1/2)$ versus the crystallographic $c/a$ ratio for a stacked triangular lattice with A-type AFM ordering [$k=(0,0,1/2)$~r.l.u.] and magnetic-moment orientations parallel to the hexagonal [0,0,1], [1,0,0], or [0,1,0] directions.}
\label{SummaryHex_0_0_0.5}
\end{figure}

Figure~\ref{SummaryHex_0_0_0.5} shows the dependence of the magnetic dipole eigenvalues $\lambda(0,0,1/2)$ for A-type AFM ordering with wavevector $k=(0,0,1/2)$ reciprocal-lattice units on a  a stacked triangular lattice as a function of the $c/a$ ratio of the crystallographic lattice from 0.5 to 3 in increments of 0.1 for all spins in a radius of 50 times the hexagonal lattice constant~$a$.  The number of spins included varied from $1.2\times10^6$ for $c/a=0.5$ to $2.0\times 10^5$ for $c/a = 3$.  As noted in the text, $\lambda(0,0,1/2)[0,0,1]$ and $\lambda(0,0,1/2)[1,0,0]/[0,1,0]$ rapidly approach the respective 2D limits of a single triangular spin layer with increasing $c/a$.  Thus for the experimental $c/a$ ratio of 1.6477 for \ems\ at 6.6~K, the eigenvalues are close to the 2D limits of $\lambda(0,0,1/2)$ for the magnetic moments aligned along [1,0,0]/[0,1,0] and [0,0,1] given by 5.517\,088 and $-11.034\,176$, respectively, as seen in Fig.~\ref{SummaryHex_0_0_0.5} and also noted in the main text.

Experimentally, in the AFM state the moments in \ems\ are aligned in the $ab$~plane as discussed in the main text.  This is consistent with Fig.~\ref{SummaryHex_0_0_0.5} and Eqs.~(\ref{Eqs:Eialpha}) which show that $ab$-plane moment alignment has a much lower energy than $c$-axis alignment.  We note that due to an oversight, the calculations in Fig.~\ref{SummaryHex_0_0_0.5} were not carried out in Ref.~\cite{Johnston2016}, which otherwise was quite comprehensive.

The values plotted in Fig.~\ref{SummaryHex_0_0_0.5} are listed in Table~\ref{Tab:AtypeTriangEvecsEvals}, which can be interpolated if desired.  The accuracy of the eigenvalues is estimated to be $\pm0.01$.

\begin{table}[h]
\caption{\label{Tab:AtypeTriangEvecsEvals} {\bf Stacked Simple Hexagonal Spin Lattices with A-type AFM order with k = (0,0,1/2).}  Eigenvalues $\lambda_{{\bf k}\alpha}$ and eigenvectors $\hat{\mu} = [\mu_x,\mu_y,\mu_z]$ in Cartesian coordinates of the magnetic dipole interaction tensor $\widehat{{\bf G}}_i({\bf k})$ in Eq.~(16c) for various values of the magnetic wavevector ${\bf k}$ in reciprocal lattice units (r.l.u.) for simple tetragonal spin lattices with {\bf collinear} magnetic moment alignments.  The most positive $\lambda_{{\bf k}\alpha}$ value(s) corresponds to the lowest energy value according to Eq.~(16d).  The Cartesian $x$, $y$ and $z$ axes are along the $a$, $b$ and $c$ axes of the tetragonal lattices, respectively. The accuracy of the values is estimated to be $\lesssim\pm 0.001$.  Also shown are the differences between the eigenvalues for different ordering axes for a given {\bf k}, which determine the anisotropy energies via Eq.~(16d).}
\begin{ruledtabular}
\begin{tabular}{cccc}
$c/a$ &		&	$\lambda_{(0,0,1/2)\alpha}$\\
	& $\alpha = [100], [010]$	& [001] & $[100]-[001]$ \\
\hline
0.5&15.301&$-$30.603&45.904\\
0.6&9.9432&$-$19.886&29.830\\
0.7&7.5810&$-$15.162&22.743\\
0.8&6.4963&$-$12.993&19.489\\
0.9&5.9848&$-$11.970&17.954\\
1.0&5.7431&$-$11.486&17.229\\
1.1&5.6246&$-$11.249&16.874\\
1.2&5.5707&$-$11.141&16.712\\
1.3&5.5417&$-$11.083&16.625\\
1.4&5.5304&$-$11.061&16.591\\
1.5&5.5219&$-$11.044&16.566\\
1.6&5.5193&$-$11.039&16.558\\
1.7&5.5181&$-$11.036&16.554\\
1.8&5.5191&$-$11.038&16.557\\
1.9&5.5184&$-$11.037&16.555\\
2.0&5.5143&$-$11.029&16.543\\
2.1&5.5185&$-$11.037&16.556\\
2.2&5.5154&$-$11.031&16.546\\
2.3&5.5183&$-$11.037&16.555\\
2.4&5.5151&$-$11.030&16.545\\
2.5&5.5207&$-$11.041&16.562\\
2.6&5.5147&$-$11.029&16.544\\
2.7&5.5172&$-$11.034&16.551\\
2.8&5.5204&$-$11.041&16.561\\
2.9&5.5152&$-$11.030&16.546\\
3.0&5.5155&$-$11.031&16.546\\
\end{tabular}
\end{ruledtabular}
\end{table}

%\newpage
\acknowledgments

We thank A. Sapkota for an x-ray Laue-diffraction measurement. The research was supported by the U.S. Department of Energy, Office of Basic Energy Sciences, Division of Materials Sciences and Engineering.  Ames Laboratory is operated for the U.S. Department of Energy by Iowa State University under Contract No.~DE-AC02-07CH11358.

%\clearpage

\end{document}